\pdfoutput=1

\documentclass[11pt,twoside,a4paper,cmspaper,final,collab]{cms-tdr}
\def\svnVersion{472372P}\def\cmsCernNoTag{CERN-EP-2018-017}\def\cmsCernDate{\today}\def\cmsMessage{Published in the Journal of High Energy Physics as \href{http://dx.doi.org/10.1007/JHEP08(2018)066}{\doi{10.1007/JHEP08(2018)066.}}}

\begin{document}\cmsNoteHeader{HIG-17-018}

\hyphenation{had-ron-i-za-tion}
\hyphenation{cal-or-i-me-ter}
\hyphenation{de-vices}
\RCS$HeadURL: svn+ssh://svn.cern.ch/reps/tdr2/papers/HIG-17-018/trunk/HIG-17-018.tex $
\RCS$Id: HIG-17-018.tex 467121 2018-07-02 18:24:30Z alverson $

\newcommand{\metHT}{\ensuremath{H_{\mathrm{T}}^\text{miss}}\xspace}
\newcommand{\metLD}{\ensuremath{L_\mathrm{D}\xspace}}

\newcommand{\mTprime}{\ensuremath{m^{'}_\mathrm{T}}}

\renewcommand{\tauh}{\ensuremath{\Pgt_{\mathrm{h}}}\xspace}
\newcommand{\Plepton}{\ensuremath{\ell}}
\newcommand{\PLepton}{\ensuremath{\ell}}
\newcommand{\Ppizero}{\ensuremath{\Pgpz}}
\newcommand{\Pnu}{\ensuremath{\PGn}}
\newcommand{\Pnut}{\ensuremath{\PGn_{\Pgt}}}
\newcommand{\APnut}{\ensuremath{\PAGn_{\Pgt}}}
\renewcommand{\Pgg}{\ensuremath{\PGg}\xspace}
\newcommand{\Pggx}{\ensuremath{\PGg^{*}}\xspace}
\newcommand{\APnu}{\ensuremath{\cPagn}\xspace}
\newcommand{\PHiggs}{\ensuremath{\PH}}
\newcommand{\Pbottom}{\ensuremath{\PQb}}
\newcommand{\Pcharm}{\ensuremath{\PQc}}
\newcommand{\APbottom}{\ensuremath{\PAQb}}
\newcommand{\ptop}{\ensuremath{\PQt}}
\newcommand{\APtop}{\ensuremath{\PAQt}}
\renewcommand{\ttbar}{\ensuremath{\ptop\APtop\text{+jets}}}
\newcommand{\ttH}{\ensuremath{\ptop\APtop\PHiggs}}
\newcommand{\tH}{\ensuremath{\ptop\PHiggs}}
\newcommand{\ttZ}{\ensuremath{\ptop\APtop\cPZ}}
\newcommand{\ttW}{\ensuremath{\ptop\APtop\PW}}
\newcommand{\ttV}{\ensuremath{\ptop\APtop\mathrm{V}}}
\newcommand{\ttWW}{\ensuremath{\ptop\APtop\PW\PW}}

\renewcommand{\ss}{\ensuremath{\text{ss}}}
\newcommand{\jet}{\ensuremath{\mathrm{j}}}
\newcommand{\bjet}{\ensuremath{\mathrm{b}}}
\newcommand{\jj}{\ensuremath{\mathrm{jj}}}
\newcommand{\vis}{\ensuremath{\text{vis}}}
\newcommand{\pass}{\ensuremath{\mathrm{p}}}

\newcommand{\MVAthadmax}{\ensuremath{\text{MVA}_{\ptop\text{had}}^{\text{max}}}}
\newcommand{\MVAHjetmax}{\ensuremath{\text{MVA}_{\PHiggs\jet}^{\text{max}}}}
\newcommand{\LR}{\ensuremath{\text{LR}}}
\newcommand{\DMVA}{\ensuremath{\mathrm{D}_{\text{MVA}}}}

\newlength\cmsTabSkip\setlength{\cmsTabSkip}{1ex}

\newcolumntype{C}[1]{>{\centering\let\newline\\\arraybackslash\hspace{0pt}}m{#1}}
\newcolumntype{L}[1]{>{\raggedright\let\newline\\\arraybackslash\hspace{0pt}}m{#1}}
\newcolumntype{R}[1]{>{\raggedleft\let\newline\\\arraybackslash\hspace{0pt}}m{#1}}

\cmsNoteHeader{HIG-17-018}
\title{Evidence for associated production of a Higgs boson with a top quark pair in final states with electrons, muons, and hadronically decaying $\PGt$ leptons at $\sqrt{s} = 13\TeV$}

\date{\today}

\abstract{
Results of a search for the standard model Higgs boson produced in association with a top quark pair ($\ttH$) in final states with electrons, muons, and hadronically decaying $\PGt$ leptons are presented. The analyzed data set corresponds to an integrated luminosity of 35.9\fbinv recorded in proton-proton collisions at $\sqrt{s} = 13\TeV$ by the CMS experiment in 2016. The sensitivity of the search is improved by using matrix element and machine learning methods to separate the signal from backgrounds. The measured signal rate amounts to $1.23^{+0.45}_{-0.43}$ times the production rate expected in the standard model, with an observed (expected) significance of $3.2\sigma$ ($2.8\sigma$), which represents evidence for $\ttH$ production in those final states. An upper limit on the signal rate of 2.1 times the standard model production rate is set at 95\% confidence level.
}

\hypersetup{
pdfauthor={CMS Collaboration},
pdftitle={Evidence for associated production of a Higgs boson with a top quark pair in final states with electrons, muons, and hadronically decaying tau leptons at sqrt(s) = 13 TeV},
pdfsubject={CMS},
pdfkeywords={CMS, physic, Higgs, top quark pair}}

\maketitle

\section{Introduction}
\label{sec:introduction}

The observation of a Higgs boson ($\PHiggs$) by the ATLAS and the CMS
experiments~\cite{Higgs-Discovery_ATLAS,Higgs-Discovery_CMS,Higgs-Discovery_CMS_long}
represents a major step towards the understanding of the mechanism for
electroweak symmetry breaking (EWSB). The current most precise measurement of the Higgs boson mass, obtained by the CMS Collaboration, is $125.26 \pm 0.21\GeV$~\cite{HIG-16-041}. The standard model (SM) makes
precise predictions for all properties of the Higgs boson, given its mass.
Within uncertainties, all measured properties of the discovered
resonance are consistent with expectations for the SM Higgs boson, corroborating the mechanism for EWSB in the SM.
In particular, the discovered particle is known to have zero spin and
positive parity \cite{ATLAS_SpinCP,HIG-14-018}. Within the present
experimental uncertainties, its
coupling to fermions is found to be proportional to the fermion mass,
as predicted by the SM. In order to confirm that the mechanism for
EWSB included in the SM is indeed realized in nature, it is important to
perform more precise measurements of the Higgs boson properties.

\begin{figure}[ht]
\centering
\includegraphics[width=0.60\textwidth]{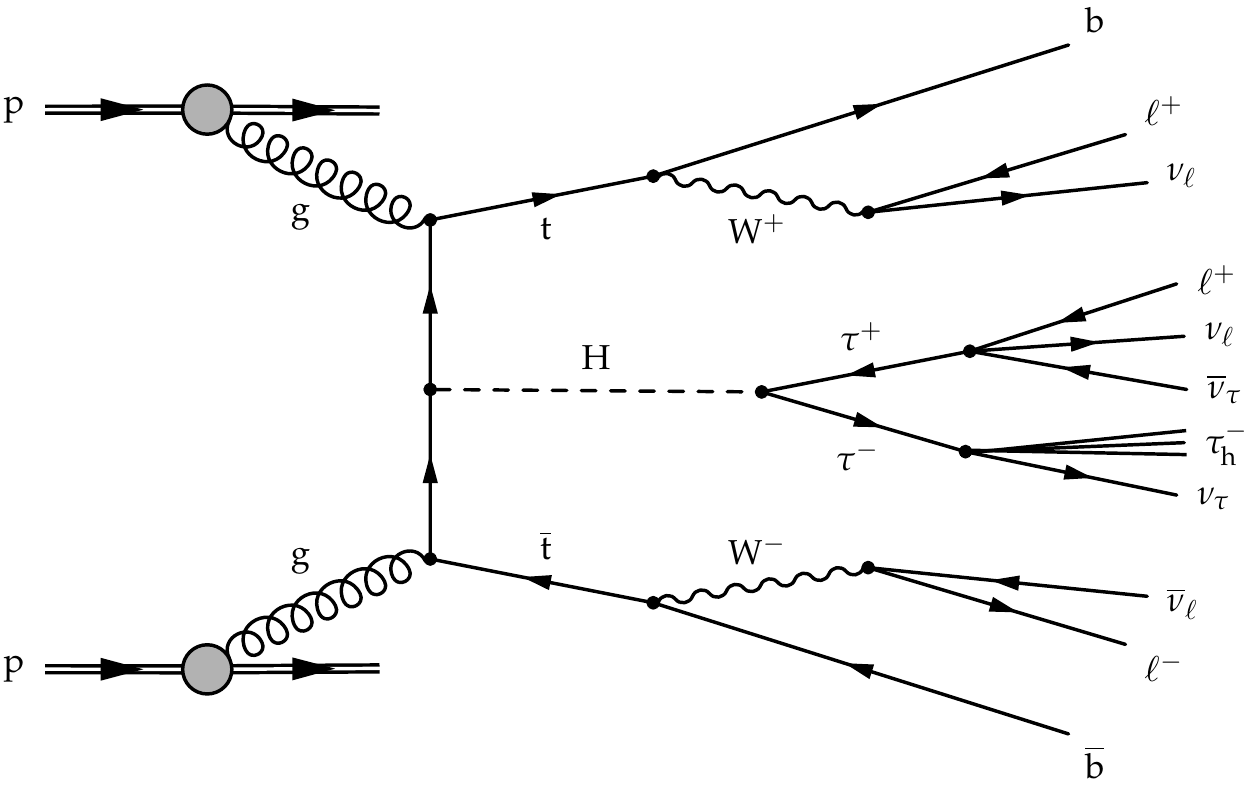}
\caption{
An example of a Feynman diagram for $\cPqt\cPaqt\PHiggs$ production, with
subsequent decay of the Higgs boson to a pair of $\Pgt$ leptons, producing a final state with two same-sign leptons and one reconstructed hadronic $\tau$ lepton decay ($\tauh$).
}
\label{fig:FeynmanDiagram_ttH_htt}
\end{figure}

The measurement of the Yukawa coupling of the Higgs boson to the top
quark, $y_{\cPqt}$, is of high phenomenological interest for several reasons.
The extraordinarily large value of the top quark mass,
compared to the masses of all other known fermions, may indicate that the top
quark plays a still-unknown special role in the EWSB mechanism.
The measurement of the rate at which Higgs bosons are produced
in association with top quark pairs ($\cPqt\cPaqt\PHiggs$ production) provides
the most precise model-independent determination of $y_{\cPqt}$.
An example of a Feynman diagram for $\cPqt\cPaqt\PHiggs$ production
in proton-proton ($\Pp\Pp$) collisions is shown in
Fig.~\ref{fig:FeynmanDiagram_ttH_htt}.
Since the rate for the gluon fusion Higgs boson production process is dominated by top quark loops,
a comparison of $y_{\cPqt}$ measured through this production channel and through the $\cPqt\cPaqt\PHiggs$ production channel
will provide powerful constraints on new physics potentially introduced into the gluon fusion process by additional loop contributions.

The associated production of a Higgs boson with a top quark pair
in $\Pp\Pp$ collisions at a center-of-mass energy of $\sqrt{s} = 8\TeV$ has been
studied in the $\PHiggs \to \Pbottom\Pbottom$ and $\PHiggs \to \Pgg\Pgg$ decay
modes as well as in multilepton final states
by the ATLAS and CMS
Collaborations~\cite{ATLAS_ttH_hbb1_8TeV, ATLAS_ttH_hbb2_8TeV, ATLAS_ttH_multilepton_with_taus_8TeV, ATLAS_ttH_and_singletopH_hgg_8TeV, CMS_ttH_hbb_multilepton_htt_hgg_8TeV, CMS_ttH_hbb_mem_8TeV}.
The final states with multiple leptons cover the decay modes
$\PHiggs \to \PW\PW$, $\PHiggs \to \cPZ\cPZ$, and  $\PHiggs \to \Pgt\Pgt$.
The ATLAS Collaboration recently reported evidence for the $\ttH$ process observed in the combination of several final states with data recorded at $\sqrt{s} = 13\TeV$~\cite{ATLAS_ttH_Hbb_13TeV, ATLAS_ttH_multilepton_13TeV}.
In this paper,
we present the results of a search for $\cPqt\cPaqt\PHiggs$ production
in multilepton final states in $\Pp\Pp$ collision data recorded with the CMS detector
at $\sqrt{s} = 13\TeV$.
The analysis is performed in six event categories,
distinguished by the number of light charged leptons (electrons and muons, generically referred to as leptons in the rest of this document)
and the number of reconstructed hadronic $\Pgt$ lepton decays in the event. We denote by the symbol $\tauh$ the system of charged and neutral hadrons produced in hadronic $\Pgt$ lepton decays.
The sensitivity of the analysis is enhanced by means of
multivariate analysis (MVA) techniques based on boosted decision trees (BDTs)~\cite{TMVA,scikit-learn}
and by matrix element method (MEM) discriminants~\cite{Kondo:1988yd, Kondo:1991dw}.

This paper is structured as follows:
the apparatus and the data samples are described in Sections~2 and 3.
Section~4 summarizes the event reconstruction.
The event selection and the background estimation are described in Sections~5 and 6.
Section~7 focuses on the signal extraction techniques.
The systematic uncertainties are discussed in Section~8.
Section~9 presents event yields, kinematic distributions, and measured properties, while the results are summarized in Section~10.
Details about the MEM computation are provided in Appendix~\ref{sec:appendix}.

\section{The CMS detector}
\label{sec:detector}

The central feature of the CMS apparatus is a superconducting solenoid of 6\unit{m} internal diameter, providing a magnetic field of 3.8\unit{T}.
A silicon pixel and strip tracker,
a lead tungstate crystal electromagnetic calorimeter (ECAL), and a brass and scintillator hadron calorimeter (HCAL),
each composed of a barrel and two endcap sections,
are positioned within the solenoid volume.
The silicon tracker measures charged particles within the pseudorapidity range $\abs{\eta}< 2.5$.
Tracks of isolated muons of transverse momentum $\pt \geq 100\GeV$ emitted at $\abs{\eta} < 1.4$ are reconstructed with an efficiency close to 100\%
and resolutions of 2.8\% in \pt and 10 (30)\mum in the transverse (longitudinal) impact parameter~\cite{TRK-11-001}.
The ECAL is a fine-grained hermetic calorimeter with quasi-projective geometry,
and is segmented into the barrel region of $\abs{\eta} < 1.48$ and in two endcaps that extend up to $\abs{\eta} < 3.0$.
The HCAL barrel and endcaps similarly cover the region $\abs{\eta} < 3.0$.
Forward calorimeters extend the coverage up to $\abs{\eta} < 5.0$.
Muons are measured and identified in the range $\abs{\eta}< 2.4$
by gas-ionization detectors embedded in the steel flux-return yoke outside the solenoid.
A two-level trigger system is used to reduce the rate of recorded
events to a level suitable for data acquisition and storage~\cite{CMS_trigger}.
The first level of the CMS trigger system, composed of custom hardware processors,
uses information from the calorimeters and muon detectors to select the most interesting events in a fixed time interval of less than 4\mus.
The high-level trigger processor farm further decreases the event rate from around 100\unit{kHz} to less than 1\unit{kHz}.
Details of the CMS detector and its performance, together with a definition of the coordinate system and the kinematic variables used in the analysis, can be found in Ref.~\cite{Chatrchyan:2008zzk}.

\section{Data samples and Monte Carlo simulation}
\label{sec:datasamples_and_MonteCarloSimulation}

The analyzed data set was collected in $\Pp\Pp$ collisions at $\sqrt{s} = 13\TeV$ in 2016
and corresponds to an integrated luminosity of 35.9\fbinv.
The events were recorded using a combination of triggers based on the presence of one, two, or three electrons and muons
or based on the presence of an electron or muon and a hadronic $\Pgt$ lepton decay.

The data are compared to signal and background estimations based on Monte Carlo (MC) simulated samples and data-driven techniques.
The main irreducible background to the analysis,
arising from the associated production of a top quark pair with one or two $\PW$ or $\cPZ$ bosons ($\ttZ$, $\ttW$, and $\ttWW$) is modeled using MC simulation.
The sum of these contributions is referred to as the $\ttV$ background.
Other relevant backgrounds that are modeled by MC simulation include
$\cPZ\Pgg$+jets, $\PW\Pgg$+jets, $\cPqt\cPaqt\Pgg$ and $\cPqt\cPaqt\Pggx$, single top,
diboson ($\PW\PW$, $\PW\cPZ$, and $\PZ\cPZ$) and triboson ($\PW\PW\PW$, $\PW\PW\PZ$, $\PW\cPZ\cPZ$, and $\PZ\cPZ\cPZ$) production,
the production of SM Higgs bosons in association with single top quarks ($\tH$),
and a few selected ``rare'' processes.
These rare processes, such as $\cPqt\cPaqt\cPqt\cPaqt$, and the production of same-sign $\PW$ boson pairs,
typically have very small cross sections, but may nevertheless yield nonnegligible background contributions.
The contribution to the signal regions from the production of SM Higgs bosons through the gluon fusion and vector boson fusion processes,
as well as their production in association with $\PW$ or $\cPZ$ bosons, is negligible.
Separate event samples are generated to simulate the production of single top quarks in association with jets, photons, and $\PW$ and $\cPZ$ bosons.
The reducible $\cPZ$+jets, $\PW$+jets and $\ttbar$ backgrounds are determined from data.
Simulated $\ttbar$ samples, produced using the leading order (LO) matrix elements implemented
in the \MGvATNLO $2.2.2$ program~\cite{MadGraph5_aMCatNLO, MadGraphLO, MadSpin},
are used solely for the purpose of validating the data-driven background estimation methods.
Samples for other background processes and for the $\ttH$ signal are generated using next-to-leading order (NLO) matrix elements implemented in the
programs \MGvATNLO and \POWHEG $v2$~\cite{MadGraphNLO,POWHEG1,POWHEG2,POWHEG3}.
The signal events are generated for a Higgs boson mass of $\MH = 125\GeV$, while a top quark mass of $\ensuremath{M_{\cPqt}} = 172.5\GeV$ is used for all simulated processes involving a top quark.
All samples are generated using the {NNPDF3.0}~\cite{NNPDF1,NNPDF2,NNPDF3} set of parton distribution functions (PDFs).
Parton shower and hadronization processes are modeled using the generator \PYTHIA $8.212$~\cite{pythia8} with the \text{CUETP8M1} tune~\cite{PYTHIA_CUETP8M1tune_CMS}.
The decays of $\Pgt$ leptons, including polarization effects, are modeled by \PYTHIA.
All samples containing top quark pairs as well as the
$\cPZ/\Pggx \to \PLepton\PLepton$ and $\PW$+jets samples are normalized
to cross sections computed at next-to-next-to-leading order
accuracy in perturbative quantum chromodynamics (pQCD)~\cite{FEWZ3,TTbarXsectionNNLO}.
The cross sections for single top quark~\cite{singleTopXsectionNNLO1,singleTopXsectionNNLO2,singleTopXsectionNNLO3} and diboson~\cite{MCFMdiBosonXsection} production are computed at NLO accuracy in pQCD.

Minimum bias events generated with \PYTHIA are overlaid on all simulated events
according to the luminosity profile of the analyzed data.
In the analyzed data set, an average of approximately 23 inelastic $\Pp\Pp$ interactions (pileup) occur per bunch crossing.

All generated events are passed through a detailed simulation of the CMS apparatus, based on \GEANTfour~\cite{geant4},
and are processed using the same version of the CMS event reconstruction software as used for data.

Small corrections are applied to simulated events as data-to-MC scale factors in order to improve the modeling of the data.
The efficiency of the triggers based on the presence of one, two, or three electrons or muons,
as well as the efficiency for electrons or muons to pass the lepton reconstruction, identification, and isolation criteria,
are measured using $\cPZ/\Pggx \to \Pe\Pe$ and $\cPZ/\Pggx \to \Pgm\Pgm$ events.
The efficiency of the triggers based on the presence of an electron or muon and a hadronic $\Pgt$ lepton decay,
the efficiency for hadronic $\Pgt$ lepton decays to pass the $\tauh$ identification criteria,
and the energy scale with which hadronic $\Pgt$ lepton decays are reconstructed, are measured using $\cPZ/\Pggx \to \Pgt\Pgt$ events~\cite{TAU-16-002}.
The $\Pbottom$ tagging efficiency and mistag rate (discussed in Section \ref{sec:jets}) are measured in $\ttbar$ and $\cPZ/\Pggx \to \PLepton\PLepton$ events~\cite{Chatrchyan:2012jua}, respectively.
The differences in the resolution of the missing transverse momentum between data and simulation are measured in $\cPZ/\Pggx \to \PLepton\PLepton$ and $\Pgg$+jets events~\cite{JME-16-004}
and corrected as described in Ref.~\cite{JME-13-003}.

\section{Event reconstruction}
\label{sec:eventReconstruction}

The information provided by all CMS subdetectors is employed by a
particle-flow (PF) algorithm~\cite{Sirunyan:2017ulk}
to identify and reconstruct individual particles in the event, namely muons, electrons, photons, and charged and neutral hadrons.
These particles are then used to reconstruct jets, $\tauh$ and the vector \pt imbalance in the event, referred to as $\ptvecmiss$,
as well as to quantify the isolation of leptons.

\subsection{Vertices}
Collision vertices are reconstructed using a deterministic annealing algorithm~\cite{Chabanat:2005zz,Fruhwirth:2007hz}.
The reconstructed vertex position is required to be compatible with the location of the LHC beam in the $x$-$y$ plane.
The reconstructed vertex with the largest value of summed physics-object $\pt^2$ is taken to be the primary $\Pp\Pp$ interaction vertex (PV).
The physics objects are the jets, clustered using the jet finding algorithm~\cite{Cacciari:2008gp,Cacciari:2011ma} with the tracks assigned to the vertex as inputs, and the associated missing transverse momentum, taken as the negative vector sum of the \pt of those jets.
Electrons, muons, and $\tauh$ candidates, which are subsequently reconstructed, are required to be compatible with originating from the selected PV.

\subsection{Electrons and muons}
\label{sec:leptons}
Electrons are reconstructed within $\abs{\eta}<2.5$ by an algorithm~\cite{EGM-13-001} that matches tracks reconstructed in the silicon tracker to energy deposits in the ECAL, without any significant energy deposit in the HCAL.
Tracks of electron candidates are reconstructed by a dedicated algorithm which accounts for the emission of bremsstrahlung photons.
The energy loss due to bremsstrahlung is determined by searching for energy deposits in the ECAL located tangentially to the track.
An MVA approach based on BDTs
is employed to distinguish electrons from hadrons mimicking an electron signature.
Observables that quantify the quality of the electron track,
the compactness of the electron cluster,
and the matching between the track momentum and direction with the sum and position of energy deposits in the ECAL are used as inputs to the BDT.
This electron identification BDT has been trained on samples of electrons and hadrons.
Additional requirements are applied in order to remove electrons originating from photon conversions~\cite{EGM-13-001}.

The identification of muons is based on
linking track segments reconstructed in the silicon tracking detector and in the muon system~\cite{Chatrchyan:2012xi} within $\abs{\eta}<2.4$.
The matching between track segments is done outside-in, starting from a track in the muon system, and inside-out,
starting from a track reconstructed in the inner detector.
If a link can be established, the track parameters are recomputed using the combination of hits in the inner and outer detectors.
Quality requirements are applied on the multiplicity of hits in the track segments, on the number of matched track segments and on the quality of the track fit~\cite{Chatrchyan:2012xi}.

Electrons and muons in signal events are expected to be isolated,
while leptons from charm ($\Pcharm$) and bottom ($\Pbottom$) quark decays as well as from in-flight decays of pions and kaons are often reconstructed near jets.
Isolated leptons are distinguished from nonisolated leptons using the scalar \pt sum over charged particles, neutral hadrons, and photons
that are reconstructed within a narrow cone centered on the lepton direction.
The size $R$ of the cone shrinks with the increasing \pt of the lepton
in order to increase the efficiency for leptons reconstructed in signal events with high hadronic activity to pass the isolation criteria.
The narrow cone size, referred to as ``mini-isolation'', has the further advantage that it reduces the effect of pileup.
Efficiency loss due to pileup is additionally reduced
by considering only those charged particles that originate from the lepton production vertex in the isolation sum.
Residual contributions of pileup to the neutral component of the isolation $I_{\Plepton}$ of the lepton are taken into account by means of so-called effective area corrections:
\begin{equation}
  I_{\Plepton} = \sum_{\text{charged}} \pt + \max \left( 0, \sum_{\text{neutrals}} \pt - \rho \, \mathcal{A} \, \left[\frac{R}{0.3}\right]^{2} \right) ,
  \label{eq:lepMiniIsolation}
\end{equation}
where $\rho$ represents the energy density of neutral particles
reconstructed within the geometric acceptance of the tracking detectors,
computed as described in Refs.~\cite{Cacciari:2008gn, Cacciari:2007fd}.
The effective area $\mathcal{A}$ is obtained from the simulation, by studying the correlation between $I_{\Plepton}$ and $\rho$,
and is determined separately for electrons and muons and in bins of $\eta$.
The size of the cone is given by:
\begin{equation}
  R =
  \left\{
    \begin{aligned}
      &0.05, &\text{if  $\pt > 200\GeV$} \\
      &10\GeV/\pt, &\text{if  $50 < \pt < 200\GeV$} \\
      &0.20, &\text{if  $\pt < 50\GeV$}
    \end{aligned}
  \right. .
  \label{eq:lepMiniIsoConeSize}
\end{equation}

Additional selection criteria are applied to discriminate leptons produced in the decays of $\PW$ bosons, $\PZ$ bosons, or $\Pgt$ leptons
from those produced in the decays of $\PB$ or light mesons.
We will refer to the former as ``prompt'' (signal) leptons and to the latter as ``nonprompt'' (background) leptons.
The separation of prompt from nonprompt leptons is performed by a BDT-based algorithm, referred to as the lepton MVA.
The following observables are used as input to the lepton MVA:
the isolation of the lepton with respect to charged and neutral particles, corrected for pileup effects;
the ratio of the \pt of the lepton to the $\pt$ of the nearest jet;
a discriminant that quantifies the probability of this jet to originate from the hadronization of a $\Pcharm$ or $\Pbottom$ quark (described in Section \ref{sec:jets});
the component of the lepton momentum perpendicular to the jet axis;
the transverse and longitudinal impact parameters of the lepton track with respect to the PV;
and the significance of the impact parameter, given by the impact parameter (in three dimensions) divided by its uncertainty, of the lepton track with respect to the PV.
The last three inputs are the \pt and $\eta$ of the lepton and an additional observable,
which improves the discrimination of prompt leptons from residual backgrounds
in which the reconstructed lepton arises from the misidentification of a light-quark or gluon jet.
For electrons, this additional observable is the output of the MVA that is used for electron identification.
For muons, it corresponds to the compatibility of track segments in the muon system with the pattern expected from muon ionization.
Inputs that require the matching of the lepton to a nearby jet are set to zero if no jet of $\pt > 10\GeV$ is reconstructed within a distance
$\Delta R = \sqrt{\smash[b]{(\eta_{\jet} - \eta_{\Plepton})^{2} + (\phi_{\jet} - \phi_{\Plepton})^{2}}} < 0.4$ from the lepton,
where $\phi$ is the azimuthal angle in radians.
Separate lepton MVAs are trained for electrons and muons, using simulated samples of prompt leptons in $\cPqt\cPaqt\PHiggs$ signal events and nonprompt leptons in $\cPqt\cPaqt$+jets background events.
Leptons selected in the signal region are required to pass a tight selection on the lepton MVA output.
Looser selection criteria for electrons and muons, referred to as the ``relaxed lepton selection'', are defined by relaxing the lepton MVA selection for the purpose of estimating the contribution of background processes
as detailed in Section~\ref{sec:backgroundEstimation}.

\subsection{Hadronic \texorpdfstring{$\Pgt$}{tau} lepton decays}
\label{sec:taus}
Hadronic $\Pgt$ lepton decays are reconstructed by the ``hadrons-plus-strips'' (HPS) algorithm~\cite{TAU-14-001} within $\abs{\eta}<2.3$.
The algorithm reconstructs individual hadronic decay modes of the $\Pgt$ lepton:
$\Pgt^{\pm} \to \Ph^{\pm}\Pnut$, $\Pgt^{\pm} \to \Ph^{\pm}\Ppizero\Pnut$, $\Pgt^{\pm} \to \Ph^{\pm}\Ppizero\Ppizero\Pnut$, and $\Pgt^{\pm} \to \Ph^{\pm}\Ph^{\mp}\Ph^{\pm}\Pnut$,
where $\Ph^{\pm}$ denotes either a charged pion or kaon.
Hadronic $\Pgt$ candidates are built by combining the charged hadrons reconstructed by the PF algorithm with neutral pions.
The neutral pions are reconstructed by clustering the photons and electrons reconstructed by the PF algorithm within rectangular strips
that are narrow in $\eta$, but wide in the $\phi$ direction,
to account for the broadening of energy deposits in the ECAL if one of the photons produced in $\Ppizero \to \Pgg\Pgg$ decays
converts within the tracking detector.
An improved version of the strip reconstruction has been developed for data analyses at 13\TeV and beyond,
replacing the one used in CMS analyses at $\sqrt{s}=7$ and 8\TeV that was based on a fixed strip size of $0.05 \times 0.20$ in $\eta$-$\phi$.
In the improved version the size of the strip is adjusted as a function of the \pt of the particles reconstructed within the strip~\cite{TAU-16-002}.

Tight isolation requirements provide the most effective way to distinguish hadronic $\Pgt$ lepton decays from a large background of light-quark and gluon jets.
The sums of scalar \pt values of charged particles and of photons are used as inputs to a BDT-based $\tauh$ identification discriminant.
Separate sums are used for charged particles that are compatible with originating from the $\tauh$ production vertex
and those that are not.
The final additions to the list of input variables are the reconstructed $\tauh$ decay mode and observables that provide sensitivity to the lifetime of the $\Pgt$ lepton.
The transverse impact parameter of the highest \pt track of the $\tauh$ candidate with respect to the PV
is used for $\tauh$ candidates reconstructed in any decay mode.
In case of $\tauh$ candidates reconstructed in the decay mode $\Pgt^{-} \to \Ph^{-}\Ph^{+}\Ph^{-}\Pnut$,
a fit of the three tracks to a common secondary vertex is attempted and the distance to the PV is used as an additional input variable to the BDT.
The isolation is computed within a cone of size $R = 0.3$, centered on the $\tauh$ direction.
Compared to the version of the HPS algorithm used by the majority of CMS analyses with hadronic $\Pgt$ lepton decays,
which use a cone of size $R = 0.5$, the size of the cone is reduced in this analysis
in order to improve the efficiency in signal events with high hadronic activity.
The BDT has been trained on samples of hadronic $\Pgt$ lepton decays in $\ttH$ signal events and jets in $\ttbar$ background events,
produced using MC simulation~\cite{TAU-16-002}.
Loose, medium, and tight working points (WPs), corresponding respectively to a 65, 55 and 45\% $\tauh$ identification efficiency and a 2, 1 and 0.5\% jet~$\to \tauh$ misidentification rate,
are defined by varying the selections on the BDT output.
The selections are adjusted as a function of the \pt of the $\tauh$ candidate
such that the $\tauh$ identification efficiency for each WP is constant as a function of \pt.
The loose WP is used for the estimation of the background due to the misreconstruction of light-quark or gluon jets as $\tauh$ candidates and is referred to as the ``relaxed $\tauh$ selection''.
Contamination from events where the reconstructed $\tauh$ originates from a misreconstructed muon or electron is reduced by requiring the reconstructed $\tauh$ not to overlap with muons or electrons passing loose selection criteria within $\Delta R < 0.3$.

\subsection{Jets}
\label{sec:jets}
Jets are reconstructed from the PF candidates
using the anti-$\kt$ algorithm~\cite{Cacciari:2008gp,Cacciari:2011ma} with a distance parameter of $0.4$,
and with the constraint that the charged particles are compatible with the selected PV.
Reconstructed jets are required not to overlap with identified electrons, muons or $\tauh$ within $\Delta R<0.4$
and to pass identification criteria that aim to reject spurious jets arising from calorimeter noise~\cite{JME-16-003}.
The energy of reconstructed jets is calibrated as a function of jet \pt and $\eta$~\cite{Chatrchyan:2011ds}.
Jet energy corrections based on the \FASTJET algorithm~\cite{Cacciari:2008gn, Cacciari:2007fd} are applied.
Jets selected for this analysis must have a $\pt > 25\GeV$ and $\abs{\eta}<2.4$.
Jets originating from the hadronization of $\Pbottom$ quarks
are identified by the ``combined secondary vertex'' algorithm~\cite{Chatrchyan:2012jua,BTV-15-001},
which exploits observables related to the long lifetime of $\Pbottom$ hadrons and to the higher particle multiplicity and mass
of $\Pbottom$ jets compared to light-quark and gluon jets.
Loose and tight b tagging criteria WPs are used, respectively associated with a mistag rate of 10 and 1\% and yielding a $\Pbottom$ jet selection efficiency of 85 and 70\%.

\subsection{Missing transverse momentum}
The $\ptvecmiss$ is calculated as the negative of the vector \pt sum of all particles reconstructed by the PF algorithm.
The magnitude of the vector is referred to as $\ptmiss$.
The $\ptmiss$ resolution and response are improved by propagating the difference between calibrated and uncalibrated jets to the $\ptmiss$
and by applying corrections that account for pileup effects, as described in Ref.~\cite{JME-13-003}.

The $\ptmiss$ is complemented by the observable $\metHT$, defined as the magnitude of the vectorial \pt sum of leptons, $\tauh$, and jets:
\begin{equation}
\metHT = \left\vert \sum_{\text{leptons}} \vec{p}_{\mathrm{T}\Plepton} + \sum_{\tauh} \vec{p}_{\mathrm{T}\Pgt} + \sum_{\text{jets}} \vec{p}_{\mathrm{Tj}} \right\vert .
\label{eq:HT}
\end{equation}
Leptons and $\tauh$ entering the sum are required to pass the relaxed selection criteria discussed in Sections \ref{sec:leptons} and \ref{sec:taus}, while the jets are required to satisfy $\pt > 25\GeV$ and $\abs{\eta} < 2.4$.
The resolution on $\metHT$ is worse compared to the resolution on $\ptmiss$.
The advantage of the observable $\metHT$ is that leptons, $\tauh$, and high \pt jets predominantly originate from the hard scattering interaction
and rarely from pileup interactions, which makes $\metHT$ less sensitive to variations in pileup conditions.

The two observables $\ptmiss$ and $\metHT$ are combined into a single linear discriminant:
\begin{equation}
\metLD = 0.6\,\ptmiss + 0.4\,\metHT ,
\label{eq:metLD}
\end{equation}
exploiting the fact that $\ptmiss$ and $\metHT$ are less correlated in events in which the reconstructed $\ptmiss$ is due to instrumental effects
compared to events with genuine $\ptmiss$
that arises from the presence of neutrinos.
The coefficients of the linear combination have been optimized to provide the best rejection against the $\cPZ$+jets background.

\section{Event selection}
\label{sec:eventSelection}

This analysis focuses on final states in which one lepton is produced in one of the top quark decays, while the additional leptons and $\tauh$ are produced in the Higgs boson or the other top quark decay.
The analysis is performed using six mutually exclusive event categories,
based on the number of reconstructed leptons and $\tauh$ candidates:
\begin{itemize}
\item one lepton and two $\tauh$ ($1\Plepton+2\tauh$),
\item two leptons with same sign of the charge (``same-sign leptons'') and zero $\tauh$ ($2\Plepton\ss$),
\item two same-sign leptons and one $\tauh$ ($2\Plepton\ss+1\tauh$),
\item three leptons and zero $\tauh$ ($3\Plepton$),
\item three leptons and one $\tauh$ ($3\Plepton+1\tauh$), and
\item four leptons ($4\Plepton$).
\end{itemize}
The categories with no $\tauh$ are mostly sensitive to the Higgs boson decay into $\PW$ or $\PZ$ bosons while the categories with at least one $\tauh$ enhance the sensitivity to the Higgs boson decay into $\tau$ leptons.
The targeted $\ttH$ decays in each category are highlighted in Tables \ref{tab:selections1} and~\ref{tab:selections2}.

Events in the $2\Plepton\ss$ and $2\Plepton\ss+1\tauh$ categories are recorded by a combination of single-lepton triggers and triggers that select events containing lepton pairs.
In the $1\Plepton+2\tauh$ category, the single-lepton triggers are complemented by triggers that select events containing an electron or muon in combination with a $\tauh$.
The efficiency to select signal events in $3\Plepton$, $3\Plepton+1\tauh$, and $4\Plepton$ categories is increased
by collecting events using a combination of single-lepton and dilepton triggers, and triggers based on the presence of three leptons.

The \pt thresholds applied in order to select the leptons in different event categories are dictated by trigger requirements.
In the $2\Plepton\ss$, $3\Plepton$, and $4\Plepton$ categories,
the lepton of highest \pt (``leading'' lepton) is required to satisfy the condition $\pt > 25\GeV$
and the lepton of second-highest \pt (``subleading'' lepton) is required to satisfy $\pt > 15\GeV$.
The third (fourth) lepton is required to have $\pt > 15 (10)\GeV$.
In the $1\Plepton+2\tauh$ category, the leading lepton is required to pass a threshold of $\pt > 25 (20)\GeV$ if it is an electron (or muon) and is restricted to be within $\abs{\eta} < 2.1$ to match the trigger requirements.
In the $2\Plepton\ss+1\tauh$ category, the leading lepton is required to satisfy $\pt > 25\GeV$,
while the subleading lepton must satisfy $\pt > 15 (10)\GeV$ if it is an electron (or muon).
In the $3\Plepton+1\tauh$ category, the leading (subleading and third) lepton is required to have $\pt>20 (10)\GeV$.

Hadronically decaying $\Pgt$ lepton candidates selected in the signal region of the $2\Plepton\ss+1\tauh$ and $3\Plepton+1\tauh$ categories
are required to pass the medium WP and must have a reconstructed $\pt > 20\GeV$.
In the $1\Plepton+2\tauh$ category, the tight WP is used instead to further reduce the dominant $\ttbar$ background.
The leading (subleading) $\tauh$ candidate in this category is required to pass a threshold of $\pt>30 (20)\GeV$.

In signal events selected in the $1\Plepton+2\tauh$ category,
the lepton predominantly originates from the leptonic decay of one of the top quarks,
while the Higgs boson decays to a pair of $\Pgt$ leptons,
which both decay hadronically.
Consequently, we require the two $\tauh$ to be of opposite sign,
the combination of signs expected for a $\tauh$ pair produced in a Higgs boson decay.
In the $2\Plepton\ss+1\tauh$ category, the sign of the reconstructed $\tauh$ is required to be opposite to that of the leptons,
while in the $3\Plepton+1\tauh$ category the sum of lepton and $\tauh$ charges is required to be zero.
Finally, the modulus of the sum of lepton charges is required to be equal to one for events selected in the $3\Plepton$,
matching the sum of charges expected in signal events.

Events selected in any category are required to contain at least one jet passing tight $\Pbottom$ tagging criteria
or at least two jets passing loose $\Pbottom$ tagging criteria.
Additional criteria on the multiplicity of jets are applied.
In the $1\Plepton+2\tauh$ and $2\Plepton\ss+1\tauh$ categories,
the presence of at least three jets, including the jets that pass the $\Pbottom$ tagging criteria, is required.
The requirement on the number of jets is tightened to at least four in the $2\Plepton\ss$ category, consistent with the higher jet multiplicity expected in this category targeting events where the $\PHiggs$ decays into $\PW\PW\to\Plepton\Pnu\cPq\cPq$.
For events selected in the $3\Plepton$, $3\Plepton+1\tauh$, and $4\Plepton$ categories,
only the presence of at least two jets is required, as those categories target events with dileptonic decay of the $\cPqt\cPaqt$ pair.

In the $2\Plepton\ss$ and $2\Plepton\ss+1\tauh$ categories,
the $\ttbar$ background is reduced significantly by requiring the two leptons to have the same sign.
Background contributions arising from events containing two leptons of opposite sign, in which the sign of one lepton is mismeasured,
are reduced by applying additional quality criteria on the charge measurement.
For electrons, the consistency of the charge measurements based on different tracking algorithms
and on hits reconstructed in either the silicon pixel detector or the combination of silicon pixel and strip detectors, is required.
For muons, the curvature of the track reconstructed based on the combination of hits in the silicon detectors and in the muon system
is required to be measured with a relative uncertainty of less than 20\%.

The probability to mismeasure the charge is significantly higher for electrons than for muons.
Background contributions to the $2\Plepton\ss$ and $2\Plepton\ss+1\tauh$ categories
that arise from $\cPZ$+jets events in which the sign of a lepton is mismeasured
are reduced by requiring events to satisfy the condition $\metLD > 30\GeV$ (applied only if both leptons are electrons in the $2\Plepton\ss+1\tauh$ category)
and vetoing events in which the mass of the electron pair is within 10\GeV of the $\cPZ$ boson mass.
In the $3\Plepton$ and $3\Plepton+1\tauh$ categories,
the background from events containing $\cPZ$ bosons ($\cPZ$+jets, $\PW\cPZ$, $\cPZ\cPZ$, and $\ttZ$)
is suppressed by requiring selected events to satisfy the condition $\metLD > 30\GeV$.
The $\cPZ$-veto is also extended to all events containing same-flavor opposite-sign (SFOS) lepton pairs
and the requirement on $\metLD$ is tightened to the condition $\metLD > 45\GeV$
for those.
For events with four or more jets the contamination from background processes with $\cPZ$ bosons is smaller and no requirement on $\metLD$ is applied.

In all categories,
events containing lepton pairs of mass less than 12\GeV are rejected,
as these events are not well modeled by the MC simulation.

In the $3\Plepton$ and $4\Plepton$ categories, events with two pairs of SFOS leptons passing loose identification criteria and with a 4-lepton invariant mass lower than 140\GeV are rejected, to avoid overlap with the dedicated $\ttH$ category from \cite{HIG-16-041}.

The event selections applied in the different categories are summarized in Tables \ref{tab:selections1} and~\ref{tab:selections2}.
Combining all the event categories and assuming the SM $\ttH$ production, 91 signal events are expected, corresponding to 0.5\% of all produced $\ttH$ events.

\begin{table}[h!t]
\topcaption{
  Event selections applied in the $2\Plepton\ss$, $2\Plepton\ss+1\tauh$, $3\Plepton$, and $3\Plepton+1\tauh$ categories.
}
\label{tab:selections1}
{\centering
\resizebox{\textwidth}{!}{
\begin{tabular}{lC{6cm}C{6cm}}
\hline
Selection & $2\Plepton\ss$ & $2\Plepton\ss+1\tauh$ \\
\hline
Targeted $\ttH$ decay & $\cPqt\to\cPqb\Plepton\Pnu$, $\cPqt\to\cPqb\cPq\cPq$, $\PHiggs\to\PW\PW\to\Plepton\Pnu\cPq\cPq$ & $\cPqt\to\cPqb\Plepton\Pnu$, $\cPqt\to\cPqb\cPq\cPq$, $\PHiggs\to\PGt\PGt\to\Plepton\tauh+\Pnu'\text{s}$ \\
[\cmsTabSkip]
Trigger & \multicolumn{2}{c}{Single- and double-lepton triggers} \\
[\cmsTabSkip]
Lepton \pt & $\pt > 25$ / $15\GeV$ & $\pt > 25$ / $15$ ($\Pe$) or $10\GeV$ ($\Pgm$) \\
Lepton $\eta$ & \multicolumn{2}{c}{$\abs{\eta} < 2.5$ ($\Pe$) or $2.4$ ($\Pgm$)} \\
$\tauh$ \pt & \NA & $\pt > 20\GeV$ \\
$\tauh$ $\eta$ & \NA & $\abs{\eta} < 2.3$ \\
Charge requirements & 2 same-sign leptons & 2 same-sign leptons \\
 & and charge quality requirements & and charge quality requirements \\
 & & $\sum\limits_{\Plepton,\tauh} q = \pm 1$ \\
[\cmsTabSkip]
Jet multiplicity & $\geq$4 jets & $\geq$3 jets \\
$\Pbottom$ tagging requirements & \multicolumn{2}{c}{$\geq$1 tight $\Pbottom$-tagged jet or $\geq$2 loose $\Pbottom$-tagged jets} \\
[\cmsTabSkip]
Missing transverse & $\metLD > 30\GeV$ & $\metLD > 30\GeV\,^{*}$ \\
momentum & & \\
[\cmsTabSkip]
Dilepton mass & \multicolumn{2}{c}{$m_{\Plepton\Plepton} > 12\GeV$ and $\abs{m_{\Pe\Pe} - m_{\PZ}} > 10\GeV\,^{*}$} \\
\hline
\end{tabular}
}

\vspace*{0.2 cm}

\resizebox{\textwidth}{!}{
\begin{tabular}{lC{6cm}C{6cm}}
\hline
Selection &  $3\Plepton$ & $3\Plepton+1\tauh$ \\
\hline
Targeted $\ttH$ decays & $\cPqt\to\cPqb\Plepton\Pnu$, $\cPqt\to\cPqb\Plepton\Pnu$, $\PHiggs\to\PW\PW\to\Plepton\Pnu\cPq\cPq$ & $\cPqt\to\cPqb\Plepton\Pnu$, $\cPqt\to\cPqb\Plepton\Pnu$, $\PHiggs\to\PGt\PGt\to\Plepton\tauh+\Pnu'\text{s}$ \\
 & $\cPqt\to\cPqb\Plepton\Pnu$, $\cPqt\to\cPqb\cPq\cPq$, $\PHiggs\to\PW\PW\to\Plepton\Pnu\Plepton\Pnu$ & \\
 & $\cPqt\to\cPqb\Plepton\Pnu$, $\cPqt\to\cPqb\cPq\cPq$, & \\
 &  $\PHiggs\to\PZ\PZ\to\Plepton\Plepton\cPq\cPq$ or $\Plepton\Plepton\Pnu\Pnu$ & \\
[\cmsTabSkip]
Trigger & \multicolumn{2}{c}{Single-, double- and triple-lepton triggers} \\
[\cmsTabSkip]
Lepton \pt & $\pt > 25$ / $15$ / $15\GeV$ & $\pt > 20$ / $10$ / $10\GeV$ \\
Lepton $\eta$ & \multicolumn{2}{c}{$\abs{\eta} < 2.5$ ($\Pe$) or $2.4$ ($\Pgm$)} \\
$\tauh$ \pt & \NA &  $\pt > 20\GeV$ \\
$\tauh$ $\eta$ & \NA & $\abs{\eta} < 2.3$ \\
Charge requirements & $\sum\limits_{\Plepton} q = \pm 1$ & $\sum\limits_{\Plepton,\tauh} q = 0$ \\
[\cmsTabSkip]
Jet multiplicity & \multicolumn{2}{c}{$\geq$2 jets} \\
$\Pbottom$ tagging requirements & \multicolumn{2}{c}{$\geq$1 tight $\Pbottom$-tagged jet or $\geq$2 loose $\Pbottom$-tagged jets} \\
[\cmsTabSkip]
Missing transverse & \multicolumn{2}{c}{No requirement if $N_{\jet} \geq 4$} \\
momentum & \multicolumn{2}{c}{$\metLD > 45\GeV\,^{\dagger}$} \\
 & \multicolumn{2}{c}{$\metLD > 30\GeV$ otherwise} \\
[\cmsTabSkip]
Dilepton mass & \multicolumn{2}{c}{$m_{\Plepton\Plepton} > 12\GeV$ and $\abs{m_{\Plepton\Plepton} - m_{\PZ}} > 10\GeV\,^{\ddagger}$} \\
Four-lepton mass & $m_{4\Plepton} > 140\GeV\,^{\S}$ & \NA  \\
\hline
\end{tabular}
}
\par}

$^{*}$ Applied only if both leptons are electrons. \\
$^{\dagger}$ If the event contains a SFOS lepton pair and $N_{\jet} \leq 3$. \\
$^{\ddagger}$ Applied to all SFOS lepton pairs. \\
$^{\S}$ Applied only if the event contains $2$ SFOS lepton pairs.
\end{table}

\begin{table}[h!t]
\topcaption{
  Event selections applied in the $1\Plepton+2\tauh$ and $4\Plepton$ categories.
}
\label{tab:selections2}
{\centering
\resizebox{\textwidth}{!}{
\begin{tabular}{lC{6cm}C{6cm}}
\hline
Selection & $1\Plepton+2\tauh$ & $4\Plepton$ \\
\hline
Targeted $\ttH$ decays & $\cPqt\to\cPqb\Plepton\Pnu$, $\cPqt\to\cPqb\cPq\cPq$, $\PHiggs\to\PGt\PGt\to\tauh\tauh+\Pnu'\text{s}$ & $\cPqt\to\cPqb\Plepton\Pnu$, $\cPqt\to\cPqb\Plepton\Pnu$, $\PHiggs\to\PW\PW\to\Plepton\Pnu\Plepton\Pnu$ \\
 & & $\cPqt\to\cPqb\Plepton\Pnu$, $\cPqt\to\cPqb\Plepton\Pnu$, \\
 & & $\PHiggs\to\PZ\PZ\to\Plepton\Plepton\cPq\cPq$ or $\Plepton\Plepton\Pnu\Pnu$ \\
[\cmsTabSkip]
Trigger & Single=lepton & Single-, double-  \\
 & and lepton+$\tauh$ triggers & and triple-lepton triggers \\
[\cmsTabSkip]
Lepton \pt & $\pt > 25$ ($\Pe$) or $20\GeV$ ($\mu$) & $\pt > 25$ / $15$ / $15$ / $10\GeV$ \\
Lepton $\eta$ & $\abs{\eta} < 2.1$ & $\abs{\eta} < 2.5$ ($\Pe$) or $2.4$ ($\Pgm$) \\
$\tauh$ \pt & $\pt>30$ / $20\GeV$ & \NA \\
$\tauh$ $\eta$ & $\abs{\eta} < 2.3$ & \NA \\
Charge requirements & $\sum\limits_{\tauh} q = 0$ and $\sum\limits_{\Plepton,\tauh} q = \pm 1$ & \NA \\
[\cmsTabSkip]
Jet multiplicity & $\geq$3 jets & $\geq$2 jets \\
$\Pbottom$ tagging requirements & \multicolumn{2}{c}{$\geq$1 tight $\Pbottom$-tagged jet or $\geq$2 loose $\Pbottom$-tagged jets} \\
[\cmsTabSkip]
Dilepton mass & $m_{\Plepton\Plepton} > 12\GeV$ & $m_{\Plepton\Plepton} > 12\GeV$ \\
 & & and $\abs{m_{\Plepton\Plepton} - m_{\PZ}} > 10\GeV\,^{\ddagger}$ \\
Four-lepton mass & \NA & $m_{4\Plepton} > 140\GeV\,^{\S}$ \\
\hline
\end{tabular}
}
\par}
$^{\dagger}$ If the event contains a SFOS lepton pair and $N_{\jet} \leq 3$. \\
$^{\ddagger}$ Applied to all SFOS lepton pairs. \\
$^{\S}$ Applied only if the event contains $2$ SFOS lepton pairs.
\end{table}

\section{Background estimation}
\label{sec:backgroundEstimation}

Contributions of background processes to the signal region (SR) of the analysis,
defined by the event selection criteria detailed in Section~\ref{sec:eventSelection},
arise from a variety of sources.
Backgrounds are categorized as being either ``reducible'' or ``irreducible''
and are estimated either from the data or modeled using the MC simulation.

A background is considered as reducible if at least one
electron or muon is due to a
nonprompt lepton (\ie, originating from the decay of a hadron) or to the misidentification of a hadron,
or if one or more $\tauh$ is due to the misidentification of a quark or gluon jet.
In the $2\Plepton\ss$ category, further sources of reducible
backgrounds arise from events containing lepton pairs of opposite
charge in which the sign of either lepton is mismeasured and from
the production of top quark pairs in association with either real or virtual
conversion photons.
The dominant reducible backgrounds, arising from the
misidentification of leptons or $\tauh$  (misidentified lepton background) or from the mismeasurement of the lepton charge
(``sign-flip'' background), are determined from data.
The procedures are described in
Sections~\ref{sec:backgroundEstimation_fakes}
and~\ref{sec:backgroundEstimation_flips}.

The background contribution arising from $\cPqt\cPaqt$ production in association with photons (``conversions'')
is mostly relevant for the $2\Plepton\ss$ and $2\Plepton\ss+1\tauh$ categories.
It is typically due to asymmetric conversions of the type $\Pgg \to \Pe^{+}\Pe^{-}$,
in which one electron or positron carries most of the energy of the photon,
while the other electron or positron is of low energy and fails to get reconstructed.
Events of this type are suppressed very effectively thanks to the electron selections used.
The small remaining background is modeled using the MC simulation.
The validity of the simulation has been verified in control regions (CRs) in data.

Irreducible background contributions are modeled using the MC simulation.
The dominant contributions are due to the production of top
quark pairs in association with $\PW$ or $\cPZ$ bosons and to the
diboson production in association with
jets, dominated by the $\PW\cPZ$ and $\PZ\cPZ$ backgrounds.
Minor contributions arise from rare SM processes such as triboson production, single top production in association with a $\cPZ$ boson, the production of two same-sign $\PW$ bosons, and $\cPqt\cPaqt\cPqt\cPaqt$ production.
Results are presented considering the $\tH$ process as a background process normalized to the SM expectation.
The SM $\tH$ rate amounts to about 5\% of the $\ttH$ one in the signal regions of this analysis.
The modeling of the data by the simulation is validated
in specific CRs, each enriched in the contribution of one of
the dominant irreducible background processes: $\ttZ$, $\ttW$, and $\PW\cPZ$+jets.

\subsection{Background from misidentified leptons and \texorpdfstring{$\tauh$}{tauh}}
\label{sec:backgroundEstimation_fakes}

The background from misidentified leptons and $\tauh$ is estimated from data by means of the
fake factor (FF) method. The method is applied to each event category
separately. It is based on selecting a sample of events
passing all selection criteria for the respective category, detailed
in Section~\ref{sec:eventSelection}, except that electrons, muons, and
$\tauh$ are required to pass the relaxed, instead of nominal, selection criteria.
We refer to these event samples as the ``application region'' (AR) of the FF method.
Events in which all leptons and $\tauh$ pass the tight
selection criteria are vetoed in order to avoid overlap with the SR.
An estimate for the contribution of the misidentified lepton
background to the SR is obtained by applying appropriately
chosen weights to the events selected in the AR.

The weights depend on the probability $f_{i}$ for a misidentified electron, muon, or
$\tauh$ that passes the relaxed selection criteria to pass the nominal
selection criteria.
For the computation of the weights, the index $i$ extends over all leptons and $\tauh$ that
pass the relaxed, but fail the nominal selection criteria.
The weights differ depending on the multiplicity of leptons and $\tauh$ passing
the relaxed selection criteria as well as on the number of those passing the nominal selection criteria, the latter being denoted by $N_{\pass}$.
For events containing a total of $2$ or $3$ objects,
the weights are given by:
\begin{equation}
\begin{aligned}
w_{2} & =
\begin{cases}
   \frac{f_{1}}{1 - f_{1}}  &  \text{if $N_{\pass} = 1$} \\[5pt]
  -\frac{f_{1} \, f_{2}}{(1 - f_{1}) \, (1 - f_{2})} \, & \text{if $N_{\pass} = 0$}
\end{cases} \\[5pt]
w_{3} & =
\begin{cases}
   \frac{f_{1}}{1 - f_{1}} \, & \text{if $N_{\pass} = 2$} \\[5pt]
  -\frac{f_{1} \, f_{2}}{(1 - f_{1}) \, (1 - f_{2})} \, &  \text{if $N_{\pass} = 1$} \\[5pt]
   \frac{f_{1} \, f_{2} \, f_{3}}{(1 - f_{1}) \, (1 - f_{2}) \, (1 - f_{3})} \, & \text{if $N_{\pass} = 0$}
\end{cases}
\end{aligned}
\label{eq:ffWeight}
\end{equation}

The sign of the weights alternates for events with different numbers of leptons and $\tauh$ candidates passing the nominal selection criteria.
The alternating sign is necessary to correctly account for the contributions
of events with different numbers of prompt leptons, nonprompt leptons, genuine $\tauh$, and hadrons
to an event sample with a given total number of reconstructed leptons and $\tauh$.
For example,
in the case of events with two leptons in the $2\Plepton\ss$ category,
the negative sign in the expression $-f_{1} \, f_{2}/[(1 - f_{1}) \, (1 - f_{2})]$ for the weight $w_{2}$
corrects for the contribution of events with two nonprompt leptons or misidentified hadrons
to the sample of events in which one lepton passes and the other one fails the nominal lepton selection criteria.
Application of the weights given by Eq.~(\ref{eq:ffWeight}) to events in the AR provides an unbiased estimate of the background contribution in the SR
arising from events with at least one nonprompt lepton or hadron misidentified as prompt lepton or $\tauh$.
A correction obtained from the MC simulation is subtracted from this estimate to account for the
contamination of the AR with irreducible backgrounds, \ie, by events in which all leptons are prompt leptons and all $\tauh$ are genuine,
and in which a prompt lepton fails the nominal lepton selection criteria or a genuine $\tauh$ fails the nominal $\tauh$ selection criteria.
The correction does not exceed 10\% of the yield in the AR in any category.

The factors $f_{i}$ are measured separately for electrons, muons, and
$\tauh$ and are parametrized as functions of \pt and $\eta$.
The CR in which the $f_{i}$ are measured is referred to as ``determination region'' (DR) of the FF method.
The $f_{i}$ for electrons and muons are measured in multijet events.
Selected events are required to contain one electron or muon passing the relaxed lepton selection criteria and at least one jet.
The data in this DR are collected with single lepton triggers,
except at low muon \pt, where the presence of an additional jet with $\pt > 40\GeV$ is required in the trigger.
Contamination from prompt leptons, primarily arising from the production of single $\PW$ and $\cPZ$ bosons in association with jets,
with a small contribution from diboson production,
is reduced by vetoing events with multiple leptons.
The residual contamination is
subtracted based on a likelihood fit, similar to the one used for measuring the $\ttH$ production rate in the SR described in Section~\ref{sec:signalExtraction},
that determines the relative contributions of different background processes with prompt leptons to the DR.
A variable closely related to the transverse mass of the electron or muon and $\ptvecmiss$,
\begin{equation}
\mTprime = \sqrt{{2 p_{\mathrm{T}\Plepton}^{\text{fix}} \, \ptmiss \, \left( 1 - \cos \Delta\phi \right)}} ,
\label{eq:mT}
\end{equation}
is used as the discriminating observable in the fit.
Here, $p_{\mathrm{T}\Plepton}^{\text{fix}}=35\GeV$ is used to reduce the correlation between $\mTprime$ and the \pt of the lepton and $\Delta\phi$ denotes the angle in the transverse plane between the lepton momentum and the $\ptvecmiss$ vector.
A complication arises from the fact that the factors $f_{i}$ are measured in multijet events,
while the dominant misidentified lepton background in the AR is due to $\ttbar$ production.
The relaxed lepton selection criteria are chosen such that the $f_{i}$ are similar for nonprompt leptons and for hadrons that are misidentified as prompt leptons
and do not differ between multijet and $\ttbar$ events.
The $f_{i}$ for $\tauh$ are measured using $\cPqt\cPaqt+$jets events
in which the two $\PW$ bosons produced in the decay of the top quark pair decay to an electron-muon pair.
The events are required to contain one electron, one muon, at least one $\tauh$ candidate passing the relaxed $\tauh$ selection,
and two or more jets, of which at least one passes the tight or at least two pass the loose $\Pbottom$ tagging criteria,
and are recorded by a combination of single-lepton triggers and triggers based on the presence of an electron-muon pair.
Contributions from other background processes are reduced by requiring the observable $\metLD$, defined by Eq.~(\ref{eq:metLD}),
to satisfy the condition $\metLD > 30\GeV$.
The contamination from background processes with genuine $\tauh$ is subtracted using the MC simulation.
Separate sets of $f_{i}$ are measured for the $\tauh$ selection criteria
applied in the $2\Plepton\ss+1\tauh$ and $3\Plepton+1\tauh$ categories
and for those applied in the $1\Plepton+2\tauh$ category.

For the $1\Plepton+2\tauh$, $2\Plepton\ss$ and $3\Plepton$ categories,
the FF method is applied as
described, whereas a modified version of the FF method is utilized in
the $2\Plepton\ss+1\tauh$ and $3\Plepton+1\tauh$ categories.
In the modified version, only the part of the misidentified lepton
background in which at least one of the reconstructed electrons or muons
is misidentified is obtained from data, relaxing only the selection criteria for electrons
and muons when defining the AR.
On the other hand, the contribution of background events that contain genuine prompt light leptons and in which
the reconstructed $\tauh$ is due to the misidentification of a quark
or gluon jet is obtained from the MC simulation,
corrected to account for the difference in the $\tauh$ misidentification probability in data and simulation.
In this way, $\ttH$ events where the reconstructed $\tauh$ is not due to a genuine $\tauh$
can be retained as signal, instead of being included in the misidentified lepton background estimate.
These events amount to $\approx$30\% of the total signal in the $2\Plepton\ss+1\tauh$ and $3\Plepton+1\tauh$ categories.

We have checked that the background due to nonprompt leptons was negligible in the $4\Plepton$ category and the FF method is therefore not used in this category.

\subsection{Sign-flip background}
\label{sec:backgroundEstimation_flips}

The sign-flip background in the $2\Plepton\ss$ and $2\Plepton\ss+1\tauh$ categories is
dominated by $\ttbar$ events with two prompt leptons in which the
sign of either prompt lepton is mismeasured.
The background is estimated from data, following a strategy similar to
the one used for the estimation of the misidentified lepton background.
The AR used to estimate the contribution of the sign-flip background to the SR
contains events passing all selection criteria of the SR,
described in Section~\ref{sec:eventSelection},
of the respective category,
except that the two leptons are required to be of opposite sign.
In the $2\Plepton\ss$ category, the sum of the probabilities to mismeasure the charge of either one of
the two leptons is then applied as an event weight.
In the $2\Plepton\ss+1\tauh$ category, only the probability to mismeasure the sign of the lepton with the same sign as the $\tauh$ is used, due to the charge requirements used for the event selection in this category.
The sign misidentification rates for electrons and muons are
measured by comparing the rates of $\cPZ/\Pggx \to \Pe\Pe$ and $\cPZ/\Pggx \to \Pgm\Pgm$ events
with leptons of the same and of opposite sign
and are parametrized as functions of lepton \pt and $\eta$.
The probability for mismeasuring the sign of electrons ranges
from 0.02\% for electrons in the barrel to 0.4\% for electrons in the endcaps, after all the object selection criteria.
The probability for mismeasuring the sign of muons is negligible in this analysis.

\section{Signal extraction}
\label{sec:signalExtraction}

The event samples selected in the SR are still dominated by backgrounds in all event categories.
The sensitivity of the statistical analysis is enhanced by extracting the signal rate by means of a maximum likelihood (ML) fit to the distribution in a discriminating observable,
except in the $4\Plepton$ category, where we resort to event counting because of the small number of events expected in this category.
In each event category, a different discriminating observable is chosen, in order to achieve the maximal separation in shape between the $\ttH$ signal and backgrounds.
The observables used for the ML fit are described in Section~\ref{sec:discriminatingObservables},
and the statistical analysis is detailed in Section~\ref{sec:statisticalAnalysis}.

\subsection{Discriminating observables}
\label{sec:discriminatingObservables}

Discriminants based on the MEM approach have been developed for the $2\Plepton\ss+1\tauh$ and $3\Plepton$ categories
to improve the separation of the $\ttH$ signal with respect to the main backgrounds.
The computation of the discriminant is based on combining the knowledge of differential theoretical cross sections
for the $\ttH$ signal and for background processes with the knowledge of the experimental resolution of the detector.
More details about their computations are provided in Appendix~\ref{sec:appendix}.

In the $2\Plepton\ss+1\tauh$ category,
a MEM discriminant $\LR(2\Plepton\ss+1\tauh)$ is directly used for the signal extraction, optimized to discriminate the $\ttH$ signal
from three types of background:
$\ttZ$ events in which the $\cPZ$ boson decays into a pair of $\Pgt$ leptons,
$\ttZ$ events in which the $\cPZ$ boson decays into a pair of electrons or muons and one lepton is misidentified as $\tauh$,
and $\ptop\APtop \to \Pbottom\Plepton\Pnu \, \APbottom\Pgt\Pnu$ events with one additional nonprompt lepton produced in either a $\Pbottom$ or a $\APbottom$ quark decay.

The discriminating observable used for the signal extraction in each of the categories $2\Plepton\ss$, $3\Plepton$, and $3\Plepton+1\tauh$ is based on the output of two BDTs.
The first BDT is trained to separate the $\ttH$ signal from the $\ttV$ background and the second to separate the signal from the $\ttbar$ background.
In the $1\Plepton+2\tauh$ category, the background is largely dominated by $\ttbar$ production and a single BDT is trained to separate the signal from this background.

The training of the BDTs is performed on simulated events. The events used for the training are not used elsewhere in the analysis.
The observables used as input to the BDTs are summarized in Table~\ref{tab:bdtInputVariables}.
The choice of input variables is optimized for each category individually,
and separate optimizations are performed for the BDT that separates the signal from the $\ttV$ background and the one that separates the signal from the $\ttbar$ background.

The input variables given in the table are defined as follows:
\begin{itemize}
\item $\Delta R(\Plepton_{1},\mathrm{j})$ ($\Delta R(\Plepton_{2},\mathrm{j})$) refers to the separation between the leading (subleading) lepton and the nearest jet;
\item $\left<\Delta R_{\jj}\right>$ refers to the average distance between jets;
\item $\Delta R_{\Pgt\Pgt}$ refers to the distance between the leading and subleading $\tauh$;
\item $N_{\jet}$ and $N_{\bjet}$ refer to the number of jets and $\Pbottom$-tagged jets of 25\GeV and $\abs{\eta}<2.4$
that do not overlap, within $R < 0.4$, with any electron, muon, or $\tauh$ passing the relaxed selection criteria;
\item $m_{\Pgt\Pgt}^{\vis}$ refers to the visible mass of the leading and subleading $\tauh$;
\item $\mT^{\Plepton 1}$ refers to the transverse mass of the leading lepton and the $\ptvecmiss$ vector, computed according to Eq.~(\ref{eq:mT});
\item $\pt^{\Plepton 1}$ ($\pt^{\Plepton 2}$, $\pt^{\Plepton 3}$) refers to the \pt of the leading (subleading, third) lepton;
\item $\pt^{\Pgt 1}$ ($\pt^{\Pgt 2}$) refers to the \pt of the leading (subleading) $\tauh$;
\item $\LR(3\Plepton)$ refers to a MEM discriminant for the $3\Plepton$ category, optimized to discriminate the $\ttH$ signal
from the two dominant irreducible background processes $\ttZ$ and $\ttW$;
\item The observable $\MVAthadmax$ quantifies the compatibility of jets with a hadronic decay of a top quark.
The compatibility is computed as the response of a BDT classifier and evaluated for each possible jet and lepton permutation, using several kinematic quantities and $\Pbottom$ tagging information as inputs.
The maximum over all those permutations is used as input to the BDT that separates the $\ttH$ signal from the $\ttbar$ background in the $2\Plepton\ss$ category;
\item The observable $\MVAHjetmax$ quantifies the compatibility of jets to originate from \linebreak
$\PHiggs\to\PW\PW^{*}$ decays in which one $\PW$ boson decays leptonically and the other to a pair of quarks.
The compatibility is computed as the response of a BDT classifier and evaluated per jet, using angular variables and jet identification variables ($\Pbottom$ tagging and quark-gluon discriminants).
The maximum over all jets is used as input to the BDT that separates the $\ttH$ signal from the $\ttV$ background in the $2\Plepton\ss$ category.
Jets that are compatible with originating from the hadronic decays of top quarks according to $\MVAthadmax$ are excluded from the computation of $\MVAHjetmax$.
\end{itemize}

\begin{table}[h!t]
\topcaption{
  Observables used as input to the BDTs that separate the $\ttH$ signal from the $\ttV$ and $\ttbar$ backgrounds
  in the $1\Plepton+2\tauh$, $2\Plepton\ss$, $3\Plepton$, and $3\Plepton+1\tauh$ categories.
}
\label{tab:bdtInputVariables}
{\centering
\begin{tabular}{lcccc}
\hline
Observable & $1\Plepton+2\tauh$ & $2\Plepton\ss$ & $3\Plepton$ & $3\Plepton+1\tauh$ \\
\hline
$\Delta R(\Plepton_{1},\mathrm{j})$ & \NA & $\surd$ & $\surd$ & $\surd$ \\
$\Delta R(\Plepton_{2},\mathrm{j})$ & \NA & $\surd$ & $\surd$ & $\surd$ \\
$\left<\Delta R_{\jj}\right>$ & $\surd$ & \NA & \NA & $\surd^{2}$ \\
$\Delta R_{\Pgt\Pgt}$ & $\surd$ & \NA & \NA & \NA \\
$\max \left( \abs{\eta^{\Plepton 1}}, \abs{\eta^{\Plepton 2}} \right)$ & \NA & $\surd$ & $\surd$ & $\surd$ \\
$\metHT$ & $\surd$ & \NA & \NA & $\surd^{2}$ \\
$N_{\jet}$ & $\surd$ & $\surd$ & $\surd$ & $\surd$ \\
$N_{\bjet}$ & $\surd$ & \NA & \NA & \NA \\
$m_{\Pgt\Pgt}^{\vis}$ & $\surd$ & \NA & \NA & \NA \\
$\mT^{\Plepton 1}$ & \NA & $\surd$ & $\surd$ & $\surd$ \\
$\pt^{\Plepton 1}$ & \NA & $\surd^{1}$ & $\surd^{1}$ & $\surd^{1}$ \\
$\pt^{\Plepton 2}$ & \NA & $\surd^{1}$ & - & - \\
$\pt^{\Plepton 3}$ & \NA & \NA & $\surd^{1}$ & $\surd^{1}$ \\
$\pt^{\Pgt 1}$ & $\surd$ & \NA & \NA & \NA \\
$\pt^{\Pgt 2}$ & $\surd$ & \NA & \NA & \NA \\
$\LR(3\Plepton)$ & \NA & \NA & $\surd^{1}$ & \NA \\
$\MVAthadmax$ & \NA & $\surd^{2}$ & \NA & \NA \\
$\MVAHjetmax$ & \NA & $\surd^{1}$ & \NA & \NA \\
\hline
\end{tabular}
\par}
$^{1}$ Used only in BDT that separates $\ttH$ signal from $\ttV$ background. \\
$^{2}$ Used only in BDT that separates $\ttH$ signal from $\ttbar$ background.
\end{table}

The outputs of the two BDTs that separate the $\ttH$ signal from the $\ttV$ and $\ttbar$ backgrounds are mapped into a single discriminant $\DMVA$
that is used as a discriminating observable for the signal extraction in the $2\Plepton\ss$, $3\Plepton$, and $3\Plepton+1\tauh$ categories.
The mapping is determined as follows.
The algorithm starts by filling two-dimensional histograms of the output of the first versus the second BDT for signal and background events.
The histograms use a fine binning. The distributions for signal and for background are smoothed using Gaussian kernels to reduce statistical fluctuations.
The ratio of signal to background event yields is computed in each bin and assigned to background events depending on the bins they fall in.
The cumulative distribution of this ratio is produced for background events and partitioned, based on its quantiles, into $N$ regions of equal background content.
The number of regions is chosen using a recursive application of the $k$-means clustering algorithm with $k=2$~\cite{Macqueen} on the two-dimensional distribution of the BDTs,
including stopping conditions limiting the statistical uncertainty in the signal and background templates.
The output of the algorithm that determines the mapping
is a partitioning of the two-dimensional plane spanned by the output of the two BDTs into $N$ regions
and an enumeration, used as a discriminant, of these regions by increasing signal-to-background ratio.
By construction, the distribution of the background is approximately flat in this discriminant,
while the distribution of the signal increases from low to high values of the discriminant.
In the $2\Plepton\ss$ and $3\Plepton$ categories, the signal extraction is performed using this discriminant in subcategories based on lepton flavor, lepton charges and b-tagging requirements.
In the $3\Plepton+1\tauh$ category, due to limited statistics in simulation, the training of the two BDTs and the two-dimensional mapping have been actually performed with an inclusive $3\Plepton$ selection, resulting in a non flat background distribution.

Events selected in the $2\Plepton\ss+1\tauh$ category are analyzed in two subcategories,
motivated by different signal-to-background ratios and different levels of signal-to-background separation provided by the MEM discriminant $\LR(2\Plepton\ss+1\tauh)$ in each of the subcategories.
The ``no-missing-jet'' subcategory contains events in which a pair of jets
compatible with originating from the hadronic decay of a $\PW$ boson is reconstructed,
which allows for a full reconstruction of the decay chain
$\ptop\APtop\PHiggs \to \Pbottom\PW \, \APbottom\PW \, \Pgt\Pgt \to \Pbottom\mathrm{jj} \, \APbottom\Plepton\Pnu \, \Plepton\APnu_{\Plepton}\Pnut \, \tauh\APnut$
in signal events,
while the ``missing-jet'' category contains events with no such pair of jets.
The full reconstruction of the decay chain improves the separation of the $\ttH$ signal from background events.
Signal events can contribute to the  ``missing-jet'' category if, for example, one of the jets produced in the $\PW$ boson decay
is outside of the \pt and $\eta$ acceptance or if it overlaps with another jet.

\subsection{Statistical analysis}
\label{sec:statisticalAnalysis}

The rate of the $\ttH$ signal $\mu$ is measured through a simultaneous ML fit to the distribution in
the discriminating observables or the number of events observed in the six event categories
$1\Plepton+2\tauh$, $2\Plepton\ss$, $2\Plepton\ss+1\tauh$, $3\Plepton$, $3\Plepton+1\tauh$, and $4\Plepton$.
The best-fit value of this parameter is denoted as $\hat{\mu}$.
A 68\% confidence interval on the parameter of interest is obtained using a maximum likelihood fit based on the profile likelihood ratio test statistic~\cite{ATL-PHYS-PUB-2011-011,HIG-11-032}.
A potential signal excess in data is quantified by calculating the corresponding $p$-value.
Upper limits on the $\ttH$ signal rate are set via the $\text{CL}_{s}$ method~\cite{Junk:1999kv,Read:2002hq}.

The nuisance parameters described in Section \ref{sec:systematicUncertainties} are treated via the frequentist paradigm,
as described in Refs.~\cite{ATL-PHYS-PUB-2011-011,HIG-11-032}.
Systematic uncertainties that affect only the normalization, but not the distribution in any discriminating observable,
are represented by $\Gamma$-function distributions if they are statistical in origin,
\eg, corresponding to the number of events observed in a control region,
and by log-normal probability density functions otherwise.
Systematic uncertainties that affect the distribution in the discriminating observables
are incorporated into the ML fit via the technique detailed in Ref.~\cite{Conway:2011in},
and represented by Gaussian probability density functions.
Nuisance parameters representing systematic uncertainties of the latter type
can also affect the normalization of the $\ttH$ signal or of the backgrounds.

\section{Systematic uncertainties}
\label{sec:systematicUncertainties}

Various imprecisely measured or simulated effects can affect the rates as well as the distributions
of the observables used for the signal extraction, described in Section~\ref{sec:signalExtraction}.
We differentiate the corresponding systematic uncertainties between experimental and theory-related sources.
The contributions of background processes that are determined from data, as described in Section~\ref{sec:backgroundEstimation},
are mostly unaffected by potential inaccuracies of the MC simulation.

The efficiency for events to pass the combination of triggers based on the presence of one, two, or three electrons or muons
is measured in bins of lepton multiplicity with an uncertainty between 1 and 3\%
using a sample of events recorded by triggers based on $\ptmiss$.
The efficiency of the trigger that selects events containing an electron or muon in combination with a $\tauh$
in the $1\Plepton+2\tauh$ category
is measured with an uncertainty of 3\% in $\cPZ/\Pggx \to \Pgt\Pgt$ events.

The efficiencies to reconstruct and identify electrons and muons are measured as a function of \pt with uncertainties ranging from 2 to 4\%
using $\cPZ/\Pggx \to \Pe\Pe$ and $\cPZ/\Pggx \to \Pgm\Pgm$ events via the tag-and-probe method discussed in Ref.~\cite{EWK-10-002}.
The $\tauh$ reconstruction and identification efficiency and the $\tauh$ energy scale
are measured with uncertainties of 5 and 3\%, respectively,
using $\cPZ/\Pggx \to \Pgt\Pgt$ events ~\cite{TAU-16-002}.

The energy scale of jets is measured using the \pt balance of jets with $\cPZ$ bosons and photons
in $\cPZ/\Pggx \to \Pe\Pe$ and $\cPZ/\Pggx \to \Pgm\Pgm$ and $\Pgg$+jets events
and the \pt balance between jets in dijet and multijet events~\cite{JME-16-001}.
The uncertainty in the jet energy scale is a few percent and depends on \pt and $\eta$.
The impact of jet energy scale uncertainties on event yields and on the distributions in kinematic observables
is evaluated by varying the jet energy corrections within their uncertainties
and propagating the effect to the final result
by recalculating all kinematic quantities, including $\ptmiss$, $\metHT$, and $\metLD$,
and reapplying all event selection criteria.

The $\Pbottom$ tagging efficiencies are measured in multijet events,
enriched in the heavy-flavor content by requiring the presence of a muon,
and in $\ttbar$ events, with uncertainties of a few percent, depending on \pt and $\eta$~\cite{BTV-15-001}.
The mistag rates for light-quark and gluon jets are measured in $\cPZ$+jets events
with an uncertainty of 5--10\% for the loose and 20--30\% for the tight $\Pbottom$ tagging criteria,
again depending on \pt and $\eta$~\cite{BTV-15-001}.

The uncertainty in the integrated luminosity amounts to 2.5\%~\cite{LUM-17-001}.

Uncertainties from theoretical sources are assigned to the $\ttV$ backgrounds and to the signal normalization.
The cross sections of the irreducible $\ttZ$, $\ttW$, and $\ttWW$ backgrounds are known with uncertainties of $^{+9.6\%}_{-11.2\%}$, $^{+12.9\%}_{-11.5\%}$, and $^{+8.1\%}_{-10.9\%}$, respectively,
from missing higher-order corrections on the perturbative expansion
and of 3.4, 4 and 3\%, respectively, from uncertainties in the PDFs and in the strong coupling constant $\alpha_{s}$~\cite{deFlorian:2016spz}.
The theoretical uncertainties in the SM expectation for the $\ttH$ signal cross section amount to $^{+5.8\%}_{-9.3\%}$ from missing higher-order corrections on the perturbative expansion
and to 3.6\% from uncertainties in the PDFs and in $\alpha_{s}$~\cite{deFlorian:2016spz}.
The effect of missing higher orders on distributions in kinematic observables
is evaluated through independent changes in the renormalization and factorization scales
by factors of $2$ and $1/2$ relative to their nominal equal values~\cite{Cacciari:2003fi,Catani:2003zt,Frederix:2011ss}.

The estimate for the misidentified lepton background, obtained from data as described in Section~\ref{sec:backgroundEstimation_fakes},
is subject to uncertainties in the factors $f_{i}$ that are used to compute the event weights in Eq.~(\ref{eq:ffWeight}).
The impact of these uncertainties is separated into effects on the normalization and on the shape of the distributions used for signal extraction.
The effect on the normalization ranges from 10 to 40\%,
depending on the multiplicity of misidentified electrons, muons, and $\tauh$, and on their \pt and $\eta$.
The uncertainties in the normalization include
the effect of statistical uncertainties in the sample used to measure the $f_{i}$,
of systematic uncertainties related to the subtraction of the prompt-lepton contamination in this sample,
and of the non-perfect agreement in simulation between distributions for misidentified lepton background and those obtained when applying the FF method.
The effect on distributions in kinematic observables is computed as follows.
In case of electrons and muons, an uncertainty band for the distributions used for signal extraction is obtained
by applying independent variations of the $f_{i}$ in different bins of \pt and $\eta$.
In case of $\tauh$, we fit the misidentification rates $f_{i}$ measured in the barrel and endcap region of the detector as function of \pt
and propagate the uncertainty in the slope of the fit to the final result, in a correlated way between all the categories with $\tauh$ candidates, with typical values around 3\%.

The uncertainty in the sign misidentification rate for electrons is propagated to the final result in a similar way.
The corresponding uncertainty in the rate of the sign-flip background amounts to $\approx$30\%.

Even though the $\PW\PZ$ production is predicted theoretically at NLO accuracy and its inclusive cross section has been measured successfully at the LHC \cite{ATLAS_WZ,CMS_WZ}, this good agreement does not translate automatically to the signal regions considered for this analysis, which require the presence of at least one $\Pbottom$-tagged jet.
A conservative 100\% uncertainty is therefore assigned to the diboson background in all categories but the $3\Plepton$ one.
The uncertainty is reduced to $\approx$40\% for the $3\Plepton$ categories from studies in a dedicated $3\Plepton$ $\PW\PZ$ CR, defined by inverting the $\PZ$ veto on the dilepton mass and the $\Pbottom$ tagging requirement. The overall uncertainty assigned to the diboson prediction in that case is estimated from the statistical uncertainty due to the limited sample size in the CR (30\%), the residual background in the CR (20\%), the uncertainties in the $\Pbottom$ tagging rate (between 10 and 40\%), and from the knowledge of PDFs and the theoretical uncertainties in the flavor composition of the jets produced in association with the electroweak bosons (up to 10\%).

An uncertainty of 50\% is assigned to the rate of other minor backgrounds.
This conservative uncertainty accounts for the fact that the small background contributions from those processes have not yet been measured at the LHC.

Among all the sources of uncertainty listed above, the ones having the largest impact on the measured $\ttH$ signal rate are related to the lepton efficiency measurement, the estimate of the misidentified lepton background and the theoretical sources affecting the normalization of the signal and irreducible backgrounds, as can be seen from Table~\ref{tab:systematics}.
The systematic uncertainties related to the lepton efficiency measurement and the estimate of the misidentified lepton background are treated as correlated between all the categories which include leptons with a given flavor.
The systematic uncertainties in the normalization of the signal and irreducible backgrounds are treated as correlated between all the categories.

\begin{table}[h!]
\topcaption{
  Summary of the main sources of systematic uncertainty and their impact on the combined measured $\ttH$ signal rate $\mu$. $\Delta\mu/\mu$ corresponds to the relative shift in signal strength obtained from varying the systematic source within its associated uncertainty.
}
\label{tab:systematics}
\centering
\begin{tabular}{lcc}
\hline
Source & Uncertainty [\%] & $\Delta\mu/\mu$ [\%]\\
\hline
\Pe, \Pgm\ selection efficiency & 2--4 & 11\\
\tauh selection efficiency & 5 & 4.5\\
$\Pbottom$ tagging efficiency & 2--15~\cite{BTV-15-001} & 6\\ [\cmsTabSkip]
Reducible background estimate & 10--40 & 11\\ [\cmsTabSkip]
Jet energy calibration & 2--15~\cite{JME-16-001} & 5\\
\tauh energy calibration & 3 & 1\\ [\cmsTabSkip]
Theoretical sources & $\approx$10  & 12\\ [\cmsTabSkip]
Integrated luminosity & 2.5 & 5\\
\hline
\end{tabular}
\end{table}

\section{Results}
\label{sec:results}

\begin{figure}
\centering
\includegraphics[width=0.49\textwidth]{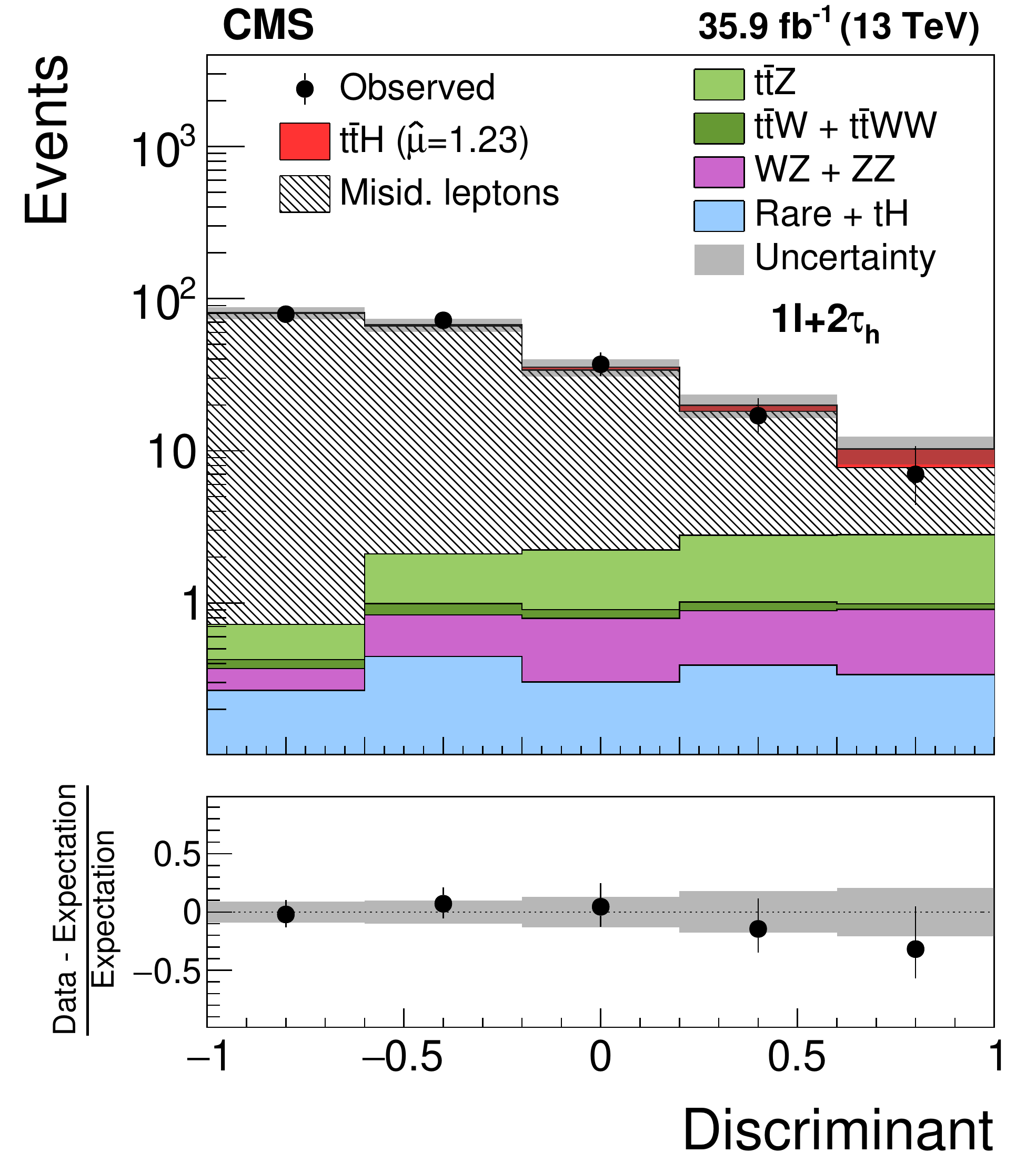}
\includegraphics[width=0.49\textwidth]{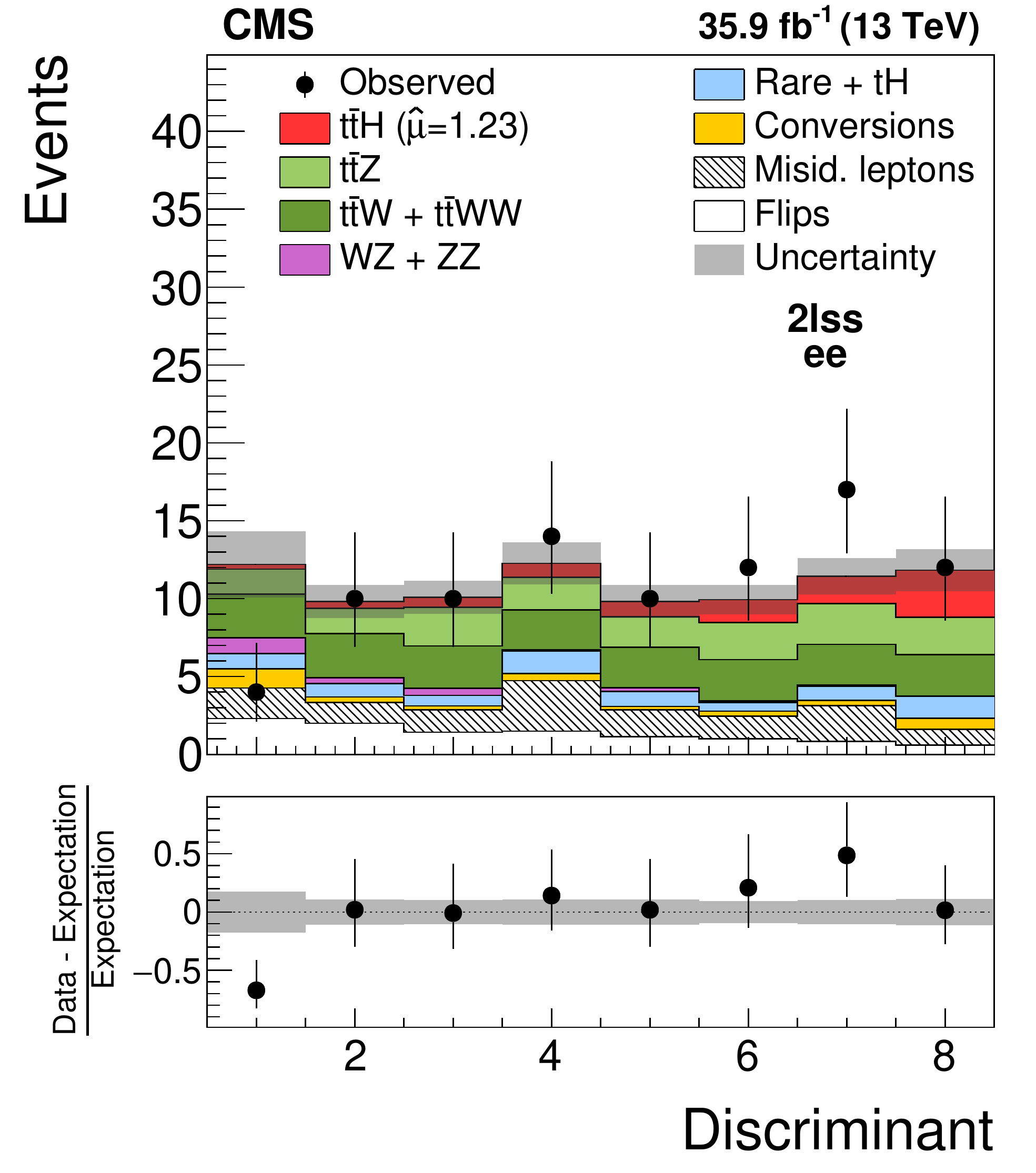}
\includegraphics[width=0.49\textwidth]{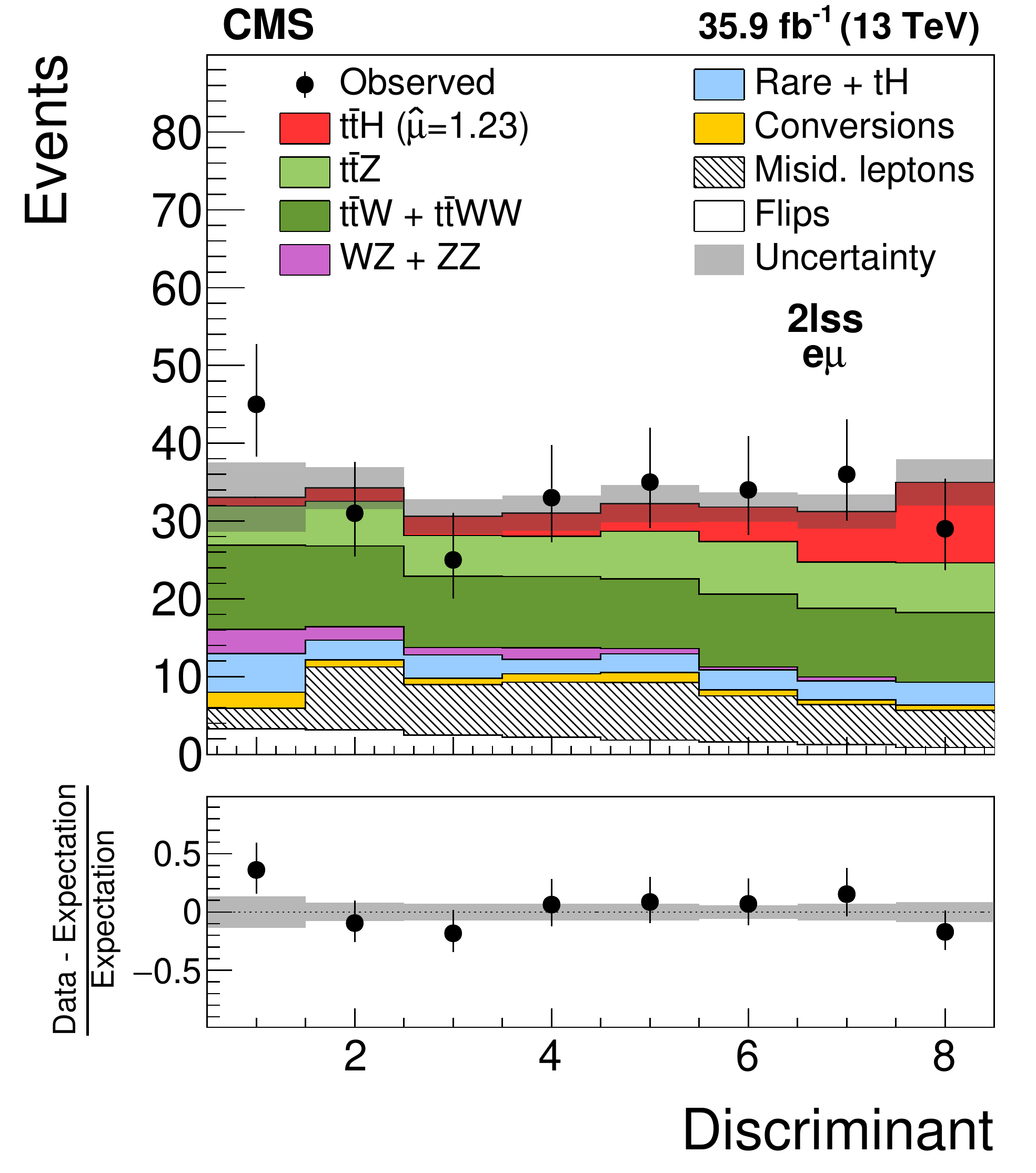}
\includegraphics[width=0.49\textwidth]{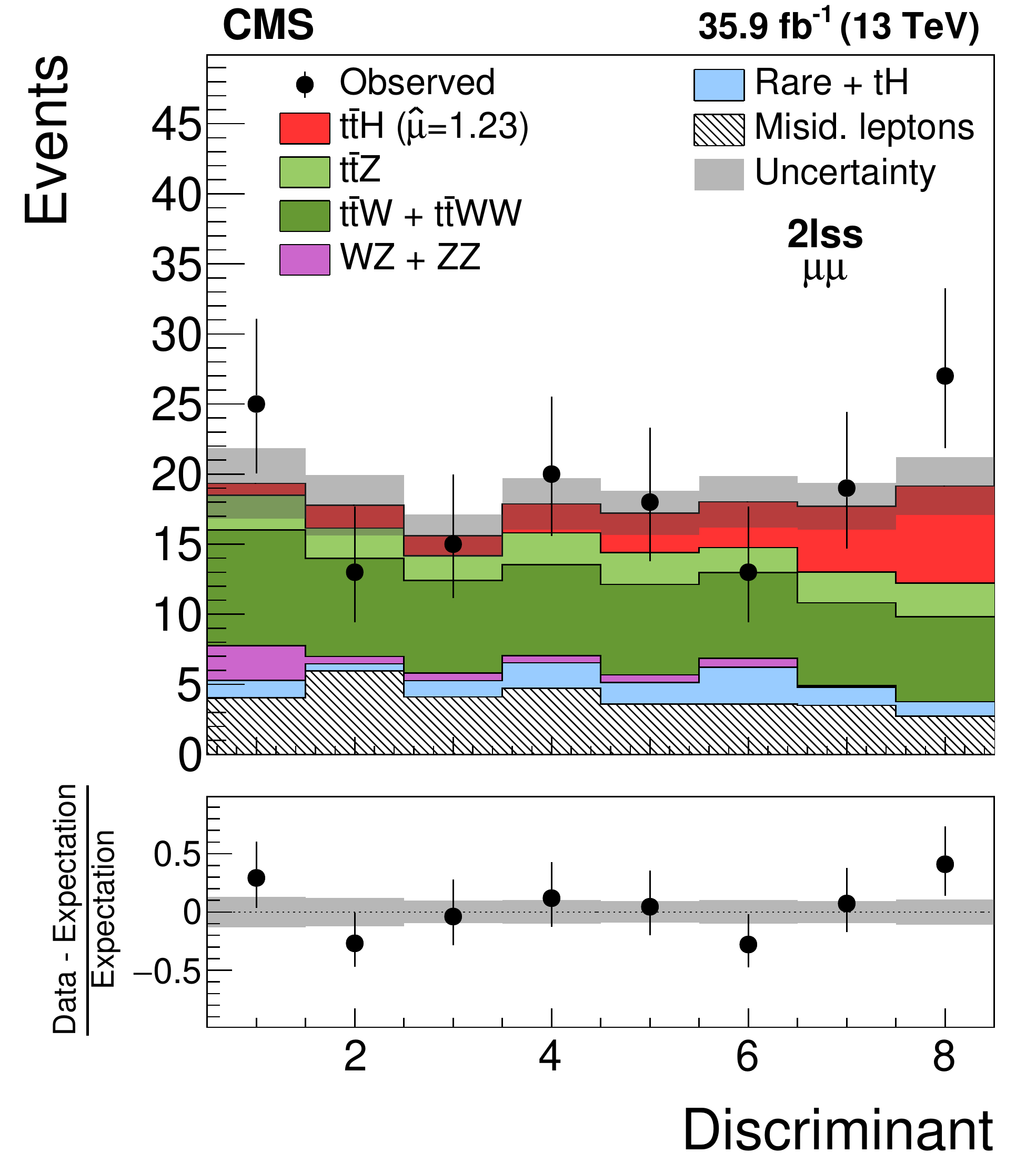}
\caption{
  Distributions in the discriminating observables used for the signal extraction in the $1\Plepton+2\tauh$ category (top left)
  and in different subcategories of the $2\Plepton\ss$ category (top right and bottom row),
  compared to the SM expectation for the
  $\ttH$ signal and for background processes.
  A BDT trained to separate the $\ttH$ signal from the $\ttbar$ background is used in the $1\Plepton+2\tauh$ category, while a $\DMVA$ variable, combining the outputs of two BDTs trained to discriminate the $\ttH$ signal from the $\ttV$ and $\ttbar$ backgrounds respectively, is used in the $2\Plepton\ss$ subcategories.
  The distributions expected for signal and background processes
  are shown for the values of nuisance parameters obtained from the combined ML fit and $\mu = \hat{\mu}=1.23$, corresponding to the best-fit value from the ML fit.
}
\label{fig:postfitPlots1}
\end{figure}

\begin{figure}
\centering
\includegraphics[width=0.49\textwidth]{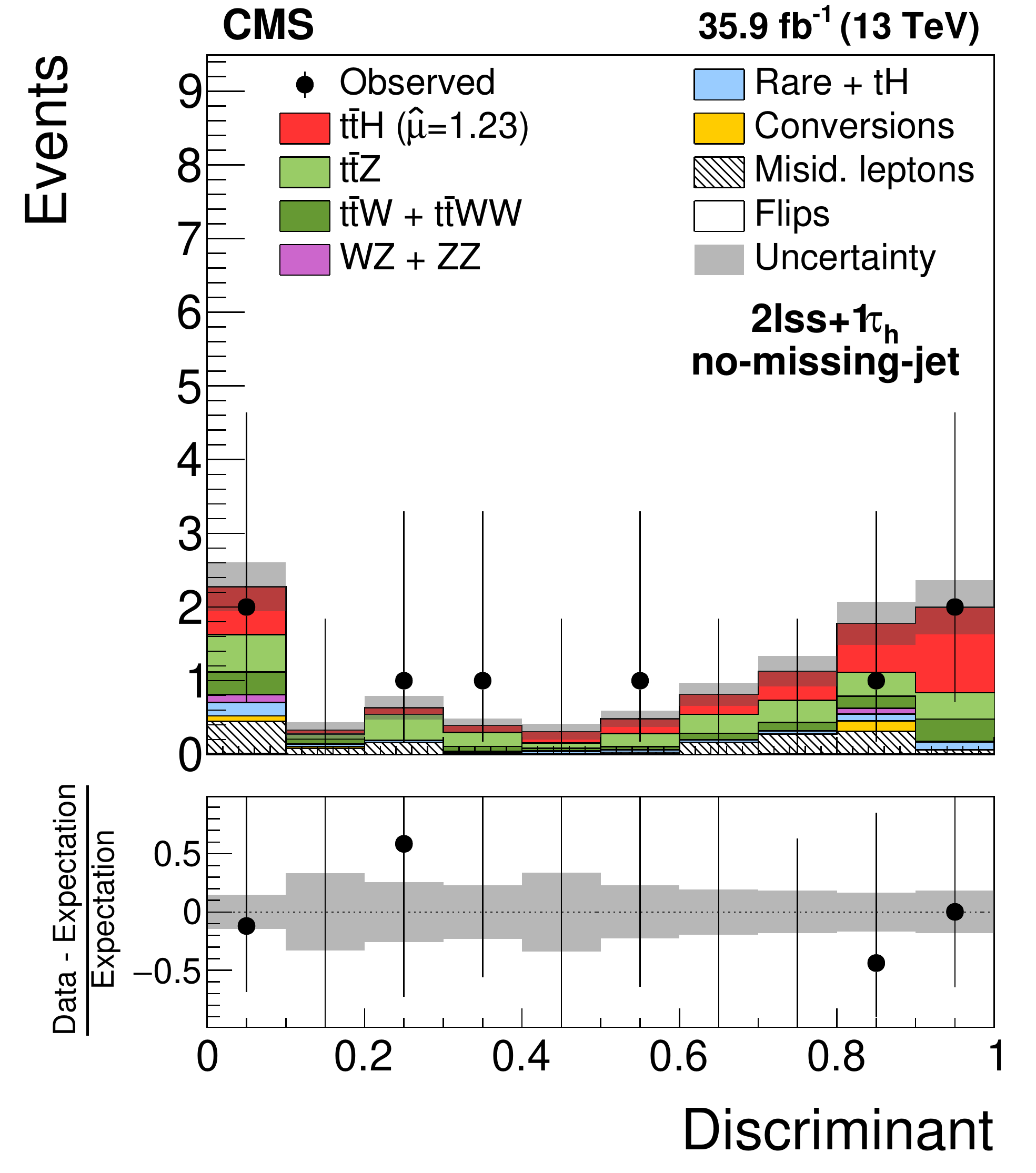}
\includegraphics[width=0.49\textwidth]{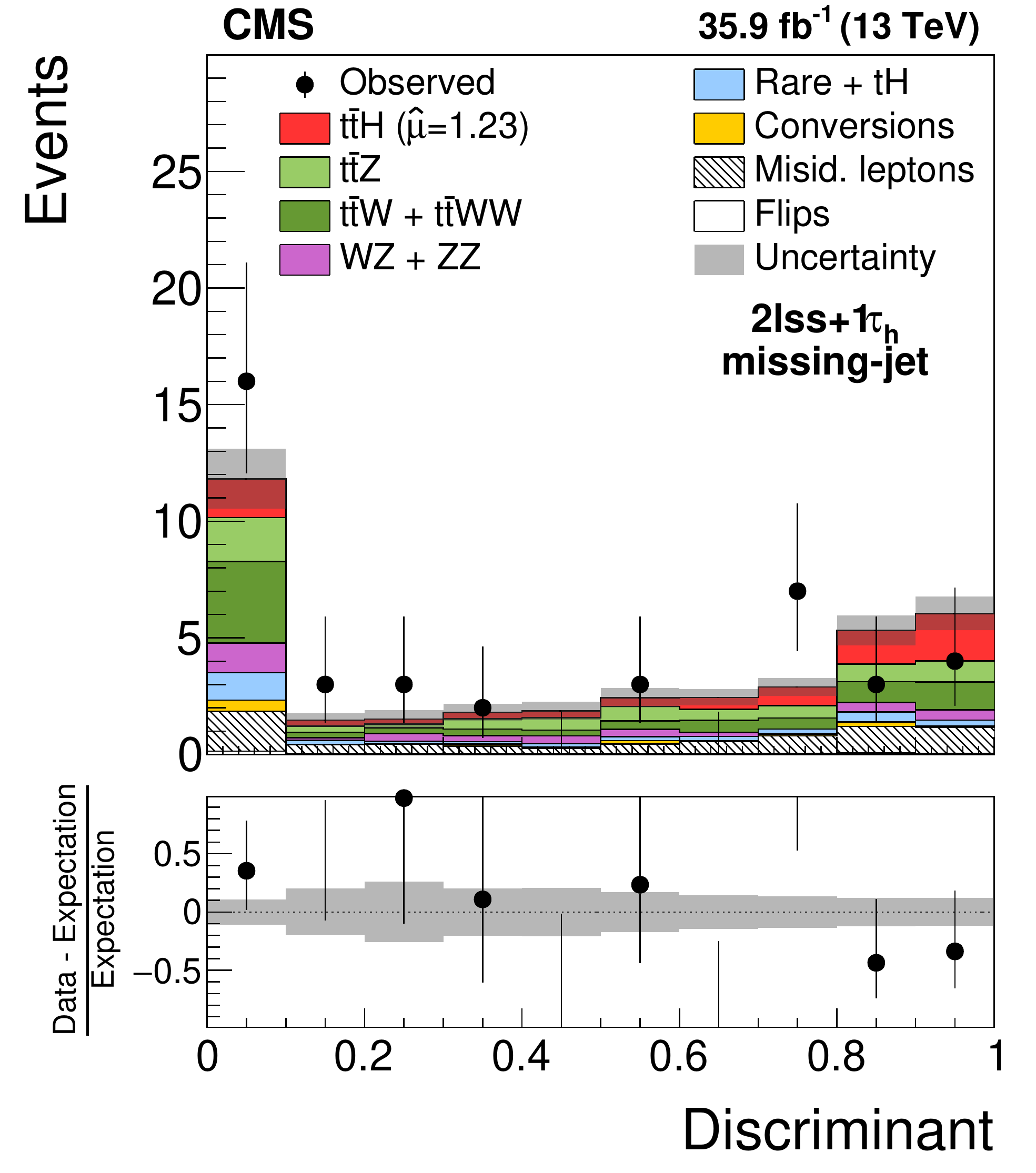}
\includegraphics[width=0.49\textwidth]{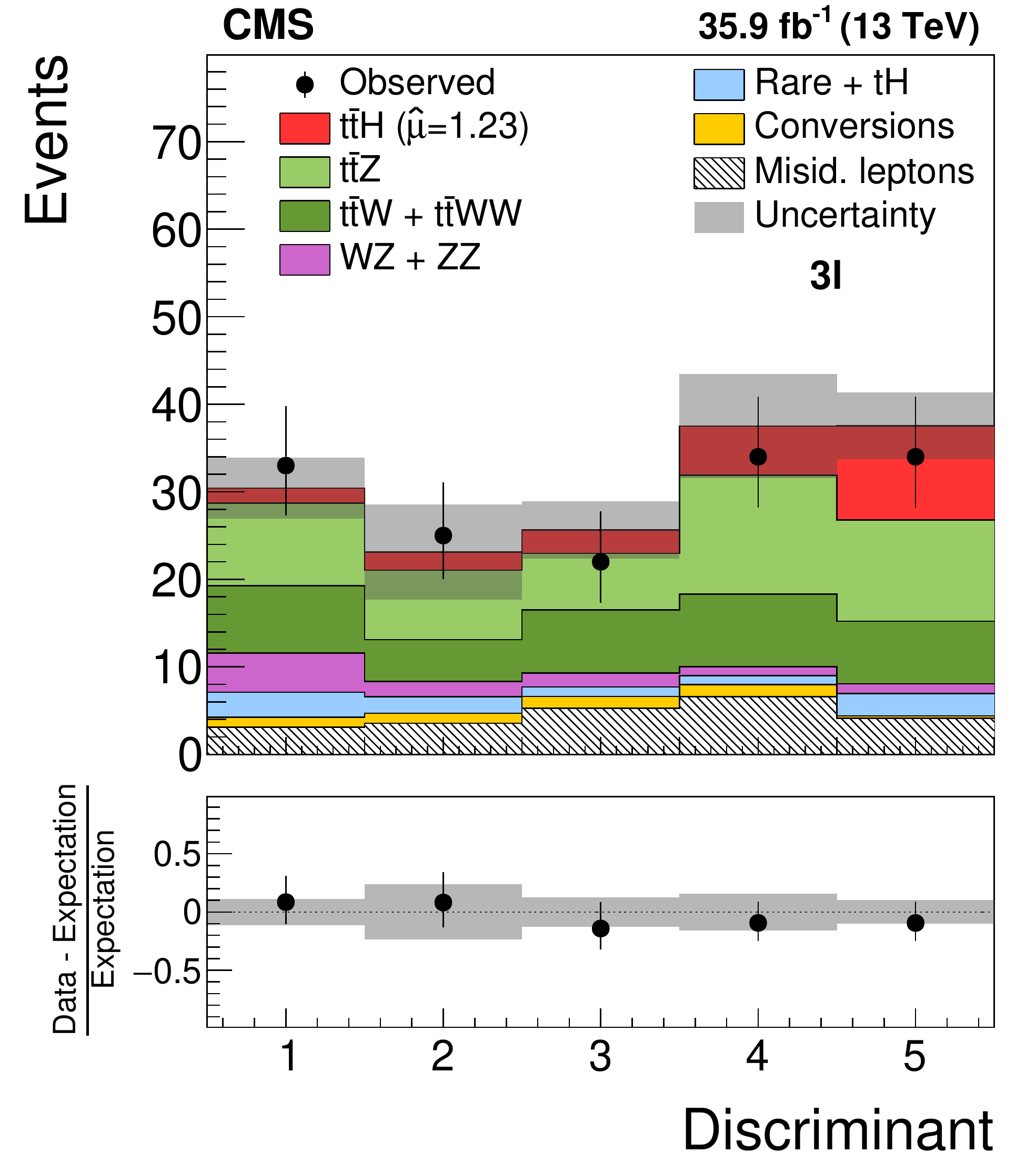}
\includegraphics[width=0.49\textwidth]{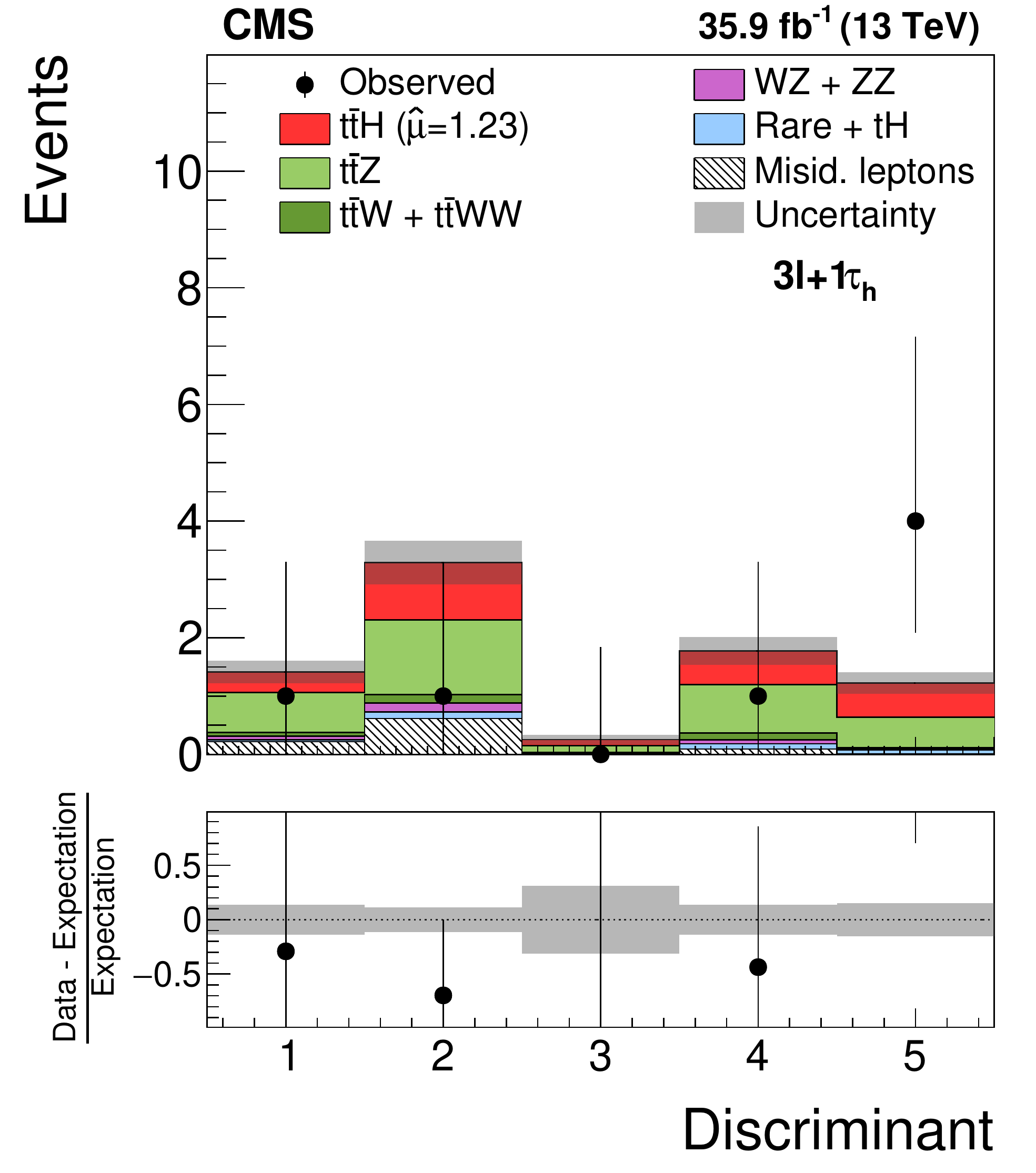}
\caption{
  Distributions in the discriminating observables used for the signal extraction in the
  ``no-missing-jet'' (top left) and ``missing-jet'' (top right) subcategories of the $2\Plepton\ss+1\tauh$ category,
  the $3\Plepton$ category (bottom left), and the $3\Plepton + 1\tauh$ category (bottom right),
  compared to the SM expectation for the
  $\ttH$ signal and for background processes.
  The MEM discriminant $\LR(2\Plepton\ss+1\tauh)$ is used in the $2\Plepton\ss+1\tauh$ subcategories, while a $\DMVA$ variable, combining the outputs of two BDTs trained to discriminate the $\ttH$ signal from the $\ttV$ and $\ttbar$ backgrounds respectively, is used in the $3\Plepton$ and $3\Plepton + 1\tauh$ categories.
  The distributions expected for signal and background processes
  are shown for the values of nuisance parameters obtained from the combined ML fit and $\mu = \hat{\mu} = 1.23$, corresponding to the best-fit value from the ML fit.
  The lowest bin of the MEM discriminant in the ``missing-jet'' subcategory of the $2\Plepton\ss+1\tauh$ category collects events
  for which the kinematics of the reconstructed objects is not compatible with the $\ttH$, $\PHiggs \to \Pgt\Pgt$ signal hypothesis.
}
\label{fig:postfitPlots2}
\end{figure}

\begin{figure}
\centering
\includegraphics[width=0.49\textwidth]{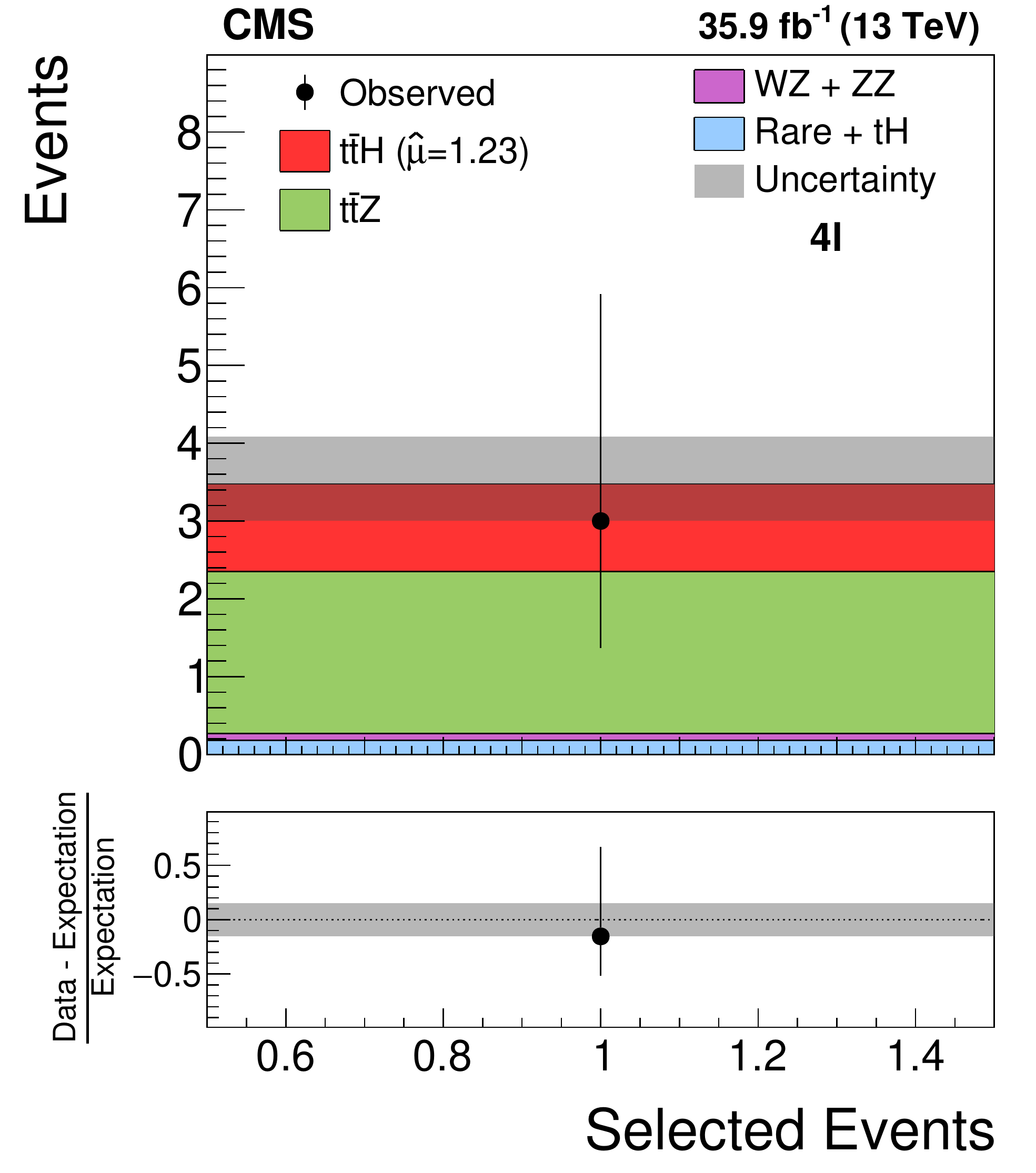}
\caption{
  Number of events observed and expected in the $4\Plepton$ category.
  The distributions expected for signal and background processes
  are shown for the values of nuisance parameters obtained from the combined ML fit and $\mu = \hat{\mu} = 1.23$, corresponding to the best-fit value from the ML fit.
}
\label{fig:postfitPlots3}
\end{figure}

\begin{figure}[h]
\centering
\includegraphics[width=0.60\textwidth]{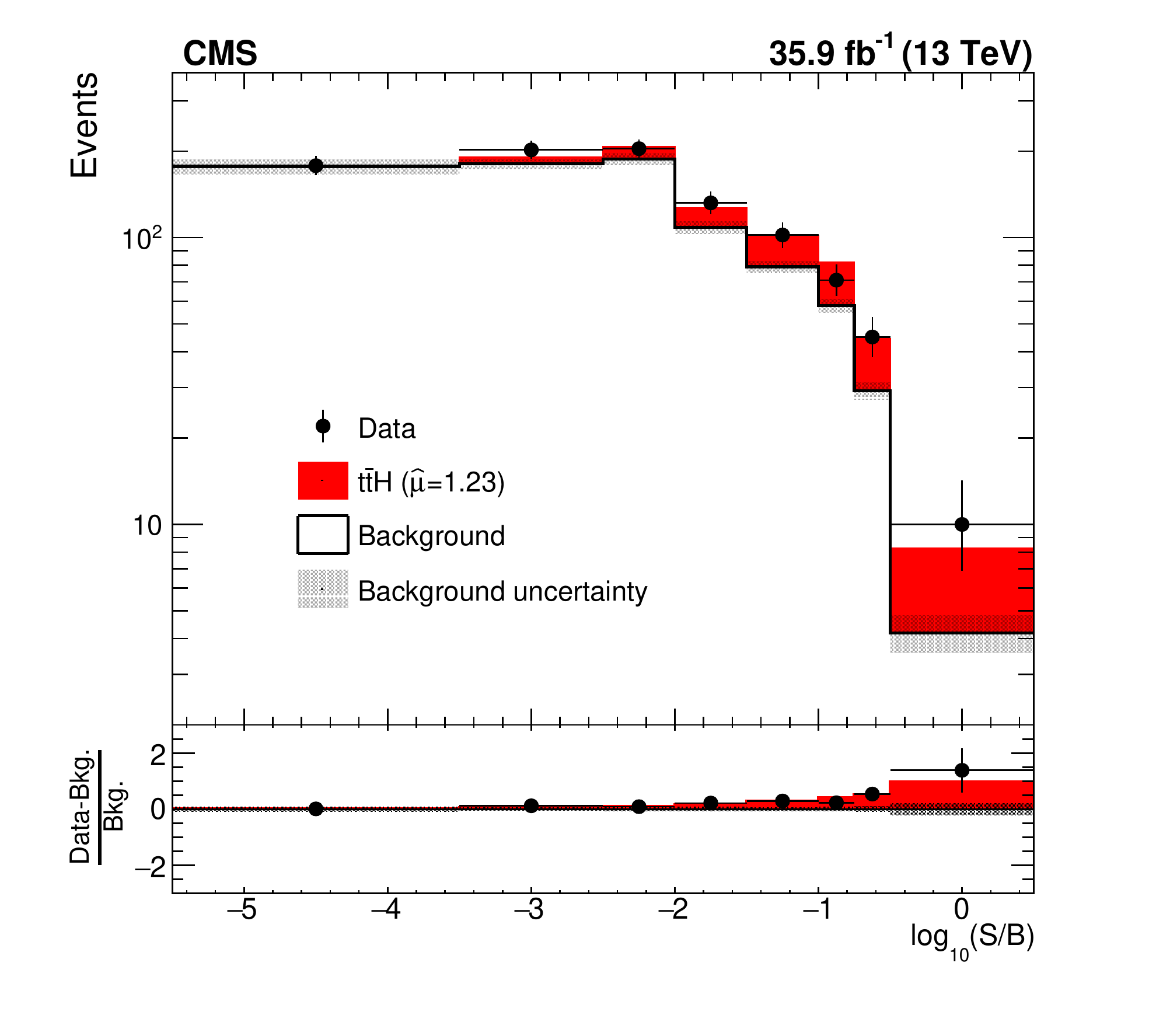}
\caption{
  Distribution of the decimal logarithm of the ratio between the expected signal and expected background in each bin of the distributions used for the signal extraction.
  The distributions expected for signal and background processes
  are shown for the values of nuisance parameters obtained from the combined ML fit and $\mu = \hat{\mu}=1.23$, corresponding to the best-fit value from the ML fit.
}
\label{fig:catPurity}
\end{figure}

The number of events observed in the different categories are compared to the SM expectations after the ML fit in Table~\ref{tab:eventYieldsPostfit}.
The event yields resulting from the fit are consistent with those predicted by the original background and signal estimates within the uncertainties described in Section~\ref{sec:systematicUncertainties}.
Most of those uncertainties are not very constrained by the ML fit, except for the uncertainty related to the background due to jets misidentified as $\tauh$ candidates.
This originates from the $1\Plepton+2\tauh$ category which is dominated by this background.
Distributions in the discriminating observables used for the signal
extraction in the different categories after the final fit are shown in Figs.~\ref{fig:postfitPlots1}--\ref{fig:postfitPlots3}.
In Fig.~\ref{fig:catPurity}, the different bins of the distributions are classified according to their expected ratio of signal (S) to background (B) events. An excess of observed events with respect to the SM backgrounds is visible in the most sensitive bins.

\begin{table}[h!]
\topcaption{
  Numbers of events selected in the different categories compared to the SM expectations
  for the $\ttH$ signal and background
  processes.
  The event yields expected for the $\ttH$ signal and for the backgrounds
  are shown for the values of nuisance parameters obtained from the combined ML fit and $\mu = \hat{\mu} = 1.23$.
  Quoted uncertainties represent the combination of statistical and systematic components.
}
\label{tab:eventYieldsPostfit}
\centering

\begin{tabular}{L{3.5cm}R{1cm}@{$ \,\,\pm\,\, $}L{1cm}R{1cm}@{$ \,\,\pm\,\, $}L{1cm}R{1cm}@{$ \,\,\pm\,\, $}L{1cm}}
\hline
Process & \multicolumn{2}{c}{$1\Plepton+2\tauh$} & \multicolumn{2}{c}{$2\Plepton\ss$} & \multicolumn{2}{c}{$2\Plepton\ss+1\tauh$} \\
\hline
$\ttH$             & $7.1$ & $2.4$ & $66.3$ & $21.0$ & \multicolumn{2}{c}{$11.6 \,\,\pm\,\, 3.5$} \\
[\cmsTabSkip]
$\ptop\APtop\cPZ/\Pggx$       & $6.3$ & $1.1$ & $80.9$ & $10.4$ & $9.2$ & $1.2$ \\
$\ttW+\ttWW$             & $0.5$ & $0.1$ & $150.0$ & $16.9$ & $9.1$ & $1.0$ \\
$\PW\cPZ+\PZ\cPZ$        & $2.1$ & $1.6$ & $16.5$ & $13.1$ & $3.9$ & $3.0$ \\
$\tH$              & $0.4$ & $0.1$ & $2.7$ & $0.2$ & $0.5$ & $0.04$ \\
Conversions        & \multicolumn{2}{c}{$<0.02$} & $12.1$ & $5.8$ & $1.4$ & $0.5$ \\
Sign flip   & \multicolumn{2}{c}{\NA} & $27.5$ & $8.0$ & $0.5$ & $0.1$ \\
Misidentified leptons               & $195.7$ & $13.6$ & $94.2$ & $21.2$ & $8.6$ & $2.1$ \\
Rare backgrounds              & $1.4$ & $0.7$ & $39.0$ & $21.2$ & \multicolumn{2}{c}{$3.1 \,\,\pm\,\, 1.5$} \\
[\cmsTabSkip]
Total expected background        & $206.3$ & $14.0$ & $423.0$ & $38.0$ & \multicolumn{2}{c}{$36.1 \,\,\pm\,\, 4.2$} \\
[\cmsTabSkip]
Observed         & \multicolumn{2}{c}{$212$} & \multicolumn{2}{c}{$507$} & \multicolumn{2}{c}{$49$} \\
\hline
\end{tabular}

\vspace*{0.2 cm}

\begin{tabular}{L{3.5cm}R{1cm}@{$ \,\,\pm\,\, $}L{1cm}R{1cm}@{$ \,\,\pm\,\, $}L{1cm}R{1cm}@{$ \,\,\pm\,\, $}L{1cm}}
\hline
Process & \multicolumn{2}{c}{$3\Plepton$} & \multicolumn{2}{c}{$3\Plepton+1\tauh$} & \multicolumn{2}{c}{$4\Plepton$} \\
\hline
$\ttH$             & $22.8$ & $7.4$ & $2.6$ & $0.9$ & \multicolumn{2}{c}{$1.1 \,\,\pm\,\, 0.4$} \\
[\cmsTabSkip]
$\ptop\APtop\cPZ/\Pggx$       & $49.0$ & $6.9$ & $3.4$ & $0.5$ & $2.1$ & $0.4$ \\
$\ttW+\ttWW$             & $35.2$ & $4.2$ & $0.4$ & $0.04$ & \multicolumn{2}{c}{$<2\times10^{-3}$} \\
$\PW\cPZ+\PZ\cPZ$        & $9.9$ & $2.4$ & $0.3$ & $0.05$ & $0.1$ & $0.1$ \\
$\tH$              & $1.2$ & $0.2$ & $0.1$ & $0.01$ & \multicolumn{2}{c}{$<4\times10^{-4}$} \\
Conversions        & $5.3$ & $2.9$ & \multicolumn{2}{c}{$<0.02$} & \multicolumn{2}{c}{$<0.02$} \\
Misidentified leptons               & $22.7$ & $6.7$ & $0.9$ & $0.2$ & \multicolumn{2}{c}{$<0.04$} \\
Rare backgrounds              & $8.2$ & $13.8$ & $0.2$ & $0.1$ & \multicolumn{2}{c}{$0.1 \,\,\pm\,\, 0.2$} \\
[\cmsTabSkip]
Total expected background        & $131.4$ & $18.2$ & $5.3$ & $0.5$ & \multicolumn{2}{c}{$2.4 \,\,\pm\,\, 0.4$} \\
[\cmsTabSkip]
Observed         & \multicolumn{2}{c}{$148$} & \multicolumn{2}{c}{$7$} & \multicolumn{2}{c}{$3$} \\
\hline
\end{tabular}

\end{table}

Upper limits on the signal rate, computed at 95\% confidence level (\CL), are given in Table~\ref{tab:limits}.
The limits are computed for separate fits of each category,
and for their combination.
The observed limit computed from the combination of
all categories amounts to $2.1$ times the SM $\ttH$ production
rate. The observed limit is compatible with the one expected if a
SM $\ttH$ signal is present at the SM predicted rate,
amounting to $1.7$ times the SM $\ttH$ production rate in the presence of a $\ttH$ signal.
In the absence of signal,
an upper limit on the signal rate of $0.8$ times the SM $\ttH$ production rate is expected.

\begin{table}
\topcaption{
  The 95\% \CL upper limits on the $\ttH$ signal rate,
  in units of the SM $\ttH$ production rate,
  obtained in each of the categories
  individually and for the combination of all six event categories.
  The observed limit is compared to the limits expected for the
  background-only hypothesis ($\mu=0$) and for the case that a $\ttH$ signal
  of SM production rate is present in the data ($\mu=1$).
  The $\pm 1$ standard deviation uncertainty intervals on the expected limits are also given in the table.
}
\label{tab:limits}

\centering
\begin{tabular}{lccc}
\hline
\multirow{2}{*}{Category} & \multirow{2}{*}{Observed limit on $\mu$} & \multicolumn{2}{c}{Expected limit} \\
 & & ($\mu=0$) & ($\mu=1$) \\
\hline
$1\Plepton+2\tauh$    & $2.7$ & $4.1^{+1.7}_{-1.4}$ & $4.8^{+2.0}_{-1.9}$ \\
$2\Plepton\ss$ & $2.8$ & $1.0^{+0.4}_{-0.2}$ & $2.0^{+0.7}_{-0.3}$ \\
$2\Plepton\ss+1\tauh$ & $2.5$ & $1.4^{+0.7}_{-0.3}$ & $2.5^{+0.9}_{-0.5}$ \\
$3\Plepton$    & $2.7$ & $1.6^{+0.8}_{-0.4}$ & $2.9^{+1.1}_{-0.4}$ \\
$3\Plepton+1\tauh$    & $4.4$ & $2.8^{+1.3}_{-0.6}$ & $4.1^{+1.5}_{-0.7}$ \\
$4\Plepton$    & $6.5$ & $4.9^{+2.8}_{-1.1}$ & $6.7^{+2.5}_{-0.8}$ \\
[\cmsTabSkip]
Combined              & $2.1$ & $0.8^{+0.3}_{-0.2}$ & $1.7^{+0.5}_{-0.5}$ \\
\hline
\end{tabular}

\end{table}

\begin{figure}[ht]
\centering
\includegraphics[width=0.70\textwidth]{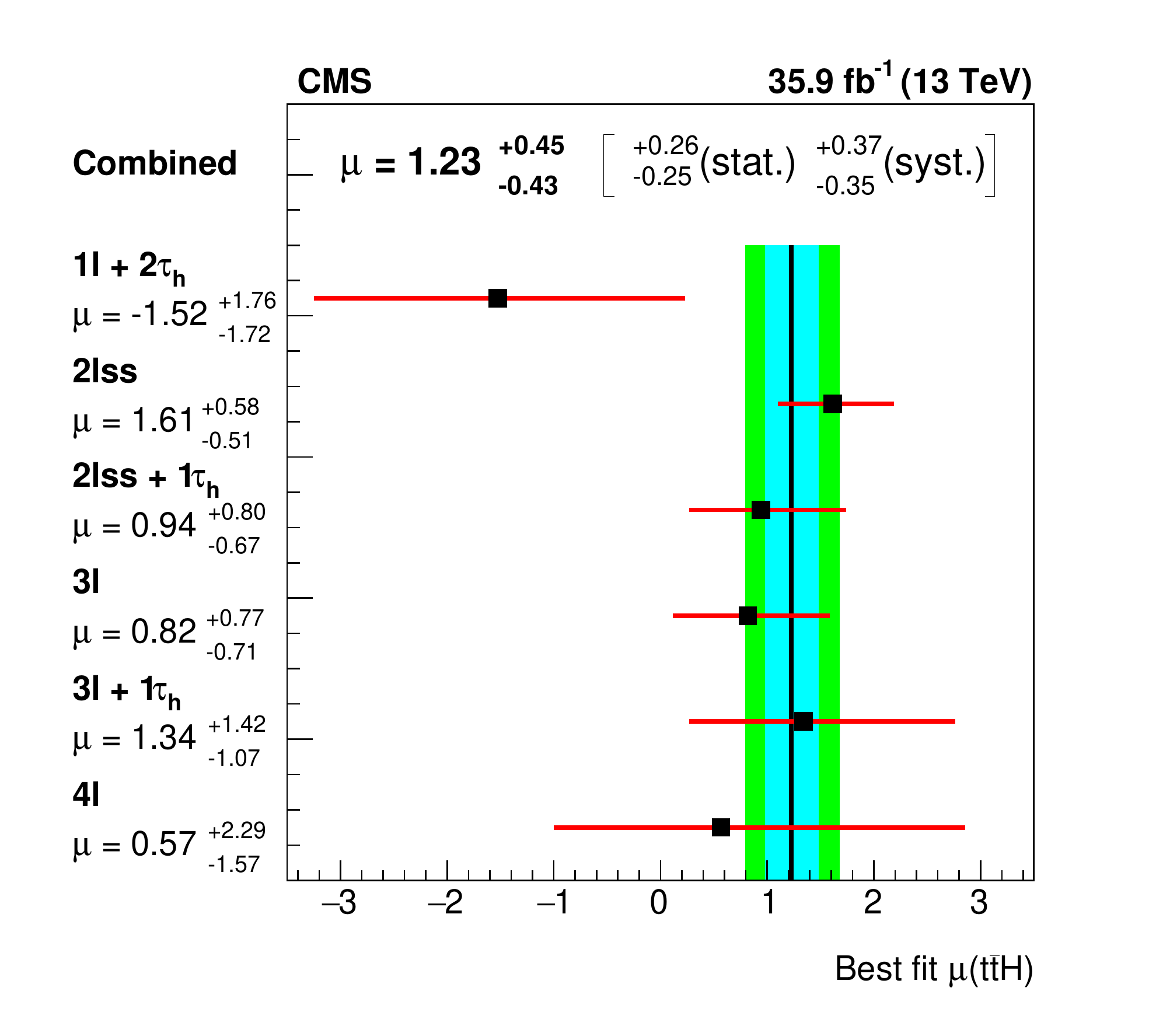}
\caption{
  Signal rates $\mu$, in units of the SM $\ttH$ production
  rate, measured in each of the categories
  individually and for the combination of all six categories.
  The blue (green) band corresponds to the statistical (total) uncertainty on the combined signal rate.
}
\label{fig:signalRates}
\end{figure}

Signal yields are extracted from a fit with $\mu$
allowed to assume different values in each category,
or constrained to assume the same value in all the categories for the combined result.
The results are shown in Fig.~\ref{fig:signalRates}.
For the combined fit,
the observed (expected) signal rate is $\mu = 1.23^{+0.45}_{-0.43}$ ($1.00^{+0.42}_{-0.38}$) times the SM $\ttH$ production rate,
with an observed (expected) significance of $3.2\sigma$ ($2.8\sigma$), which represents evidence for $\ttH$ production in those final states.
While the categories $2\Plepton\ss$, $3\Plepton$ and $4\Plepton$ are mostly sensitive to the $\ttH$ signal in the $\PHiggs\to\PW\PW$ and $\PHiggs\to\PZ\PZ$ decay modes, the $1\Plepton+2\tauh$, $2\Plepton\ss+1\tauh$ and $3\Plepton+1\tauh$ categories enhance the sensitivity to the $\PHiggs\to\Pgt\Pgt$ decay mode.
The distributions in the discriminating observables are very similar for $\ttH$ signal events with a $\PHiggs$ boson decaying into $\PW$ bosons, $\PZ$ bosons, and $\Pgt$ leptons, however, causing a large anti-correlation between the corresponding signal rates.
Denoting the sum of $\PHiggs\to\PW\PW$ and $\PHiggs\to\PZ\PZ$ decay modes by $\PHiggs\to\mathrm{VV}$ and performing a two-parameter simultaneous fit for the signal rates $\mu(\ttH,\PHiggs\to\mathrm{VV})$ and $\mu(\ttH,$ $\PHiggs\to\Pgt\Pgt)$, we obtain $\mu(\ttH,\PHiggs\to\mathrm{VV}) = 1.69^{+0.68}_{-0.61}$ and $\mu(\ttH,$ $\PHiggs\to\Pgt\Pgt) = 0.15^{+1.07}_{-0.91}$. The expected anti-correlation between the two measured signal strengths has been explicitly checked and is associated with a correlation factor of $-0.45$.

As a cross check, the analysis is repeated with the $\ttZ$ and $\ttW(W)$ backgrounds
kept freely floating in the ML fit.
Control regions enriched in the contributions of these backgrounds
are added to the fit to constrain them.
The $\ttZ$-enriched control region is defined from the $3\Plepton$ signal region by inverting the $\PZ$ boson veto on the invariant mass of SFOS lepton pairs.
The $\ttW$-enriched control region is defined from the $2\Plepton\ss$ signal region but changing the jet multiplicity requirement to consider events with exactly three jets.
The signal rate obtained from this fit is
$\mu = 1.04^{+0.50}_{-0.36}$ ($1.00^{+0.42}_{-0.38}$) times the SM $\ttH$ production rate,
with an observed (expected) significance of $2.7\sigma$ ($2.7\sigma$).

\clearpage

\section{Summary}
\label{sec:summary}

A search has been presented for the associated production of a Higgs boson with a top quark pair in final states with electrons, muons, and hadronically decaying $\Pgt$ leptons ($\tauh$).
The analyzed data set corresponds to an integrated luminosity of 35.9\fbinv of $\Pp\Pp$ collision data recorded by the CMS experiment at $\sqrt{s} = 13\TeV$.
The analysis is performed in six mutually exclusive event categories, based on different lepton and $\tauh$ multiplicity requirements.
The sensitivity of the analysis is enhanced by using multivariate analysis techniques based on boosted decision trees and on the matrix element method.
The results of the analysis are in agreement with the standard model (SM) expectation.
The measured signal rate amounts to $1.23^{+0.45}_{-0.43}$ times the SM $\ttH$ production rate, with an observed (expected) significance of $3.2\sigma$ ($2.8\sigma$), which represents evidence for $\ttH$ production in those final states. An upper limit on the signal rate of $2.1$ times the SM $\ttH$ production rate is set at 95\% confidence level, for an expected limit of $1.7$ times the SM production rate in the presence of a $\ttH$ signal.

\begin{acknowledgments}
We congratulate our colleagues in the CERN accelerator departments for the excellent performance of the LHC and thank the technical and administrative staffs at CERN and at other CMS institutes for their contributions to the success of the CMS effort. In addition, we gratefully acknowledge the computing centers and personnel of the Worldwide LHC Computing Grid for delivering so effectively the computing infrastructure essential to our analyses. Finally, we acknowledge the enduring support for the construction and operation of the LHC and the CMS detector provided by the following funding agencies: BMWFW and FWF (Austria); FNRS and FWO (Belgium); CNPq, CAPES, FAPERJ, and FAPESP (Brazil); MES (Bulgaria); CERN; CAS, MoST, and NSFC (China); COLCIENCIAS (Colombia); MSES and CSF (Croatia); RPF (Cyprus); SENESCYT (Ecuador); MoER, ERC IUT, and ERDF (Estonia); Academy of Finland, MEC, and HIP (Finland); CEA and CNRS/IN2P3 (France); BMBF, DFG, and HGF (Germany); GSRT (Greece); NKFIA (Hungary); DAE and DST (India); IPM (Iran); SFI (Ireland); INFN (Italy); MSIP and NRF (Republic of Korea); LAS (Lithuania); MOE and UM (Malaysia); BUAP, CINVESTAV, CONACYT, LNS, SEP, and UASLP-FAI (Mexico); MBIE (New Zealand); PAEC (Pakistan); MSHE and NSC (Poland); FCT (Portugal); JINR (Dubna); MON, RosAtom, RAS, RFBR and RAEP (Russia); MESTD (Serbia); SEIDI, CPAN, PCTI and FEDER (Spain); Swiss Funding Agencies (Switzerland); MST (Taipei); ThEPCenter, IPST, STAR, and NSTDA (Thailand); TUBITAK and TAEK (Turkey); NASU and SFFR (Ukraine); STFC (United Kingdom); DOE and NSF (USA).

\hyphenation{Rachada-pisek} Individuals have received support from the Marie-Curie program and the European Research Council and Horizon 2020 Grant, contract No. 675440 (European Union); the Leventis Foundation; the A. P. Sloan Foundation; the Alexander von Humboldt Foundation; the Belgian Federal Science Policy Office; the Fonds pour la Formation \`a la Recherche dans l'Industrie et dans l'Agriculture (FRIA-Belgium); the Agentschap voor Innovatie door Wetenschap en Technologie (IWT-Belgium); the F.R.S.-FNRS and FWO (Belgium) under the ``Excellence of Science - EOS" - be.h project n. 30820817; the Ministry of Education, Youth and Sports (MEYS) of the Czech Republic; the Lend\"ulet ("Momentum") Program and the J\'anos Bolyai Research Scholarship of the Hungarian Academy of Sciences, the New National Excellence Program \'UNKP, the NKFIA research grants 123842, 123959, 124845, 124850 and 125105 (Hungary); the Council of Science and Industrial Research, India; the HOMING PLUS program of the Foundation for Polish Science, cofinanced from European Union, Regional Development Fund, the Mobility Plus program of the Ministry of Science and Higher Education, the National Science Center (Poland), contracts Harmonia 2014/14/M/ST2/00428, Opus 2014/13/B/ST2/02543, 2014/15/B/ST2/03998, and 2015/19/B/ST2/02861, Sonata-bis 2012/07/E/ST2/01406; the National Priorities Research Program by Qatar National Research Fund; the Programa Estatal de Fomento de la Investigaci{\'o}n Cient{\'i}fica y T{\'e}cnica de Excelencia Mar\'{\i}a de Maeztu, grant MDM-2015-0509 and the Programa Severo Ochoa del Principado de Asturias; the Thalis and Aristeia programs cofinanced by EU-ESF and the Greek NSRF; the Rachadapisek Sompot Fund for Postdoctoral Fellowship, Chulalongkorn University and the Chulalongkorn Academic into Its 2nd Century Project Advancement Project (Thailand); the Welch Foundation, contract C-1845; and the Weston Havens Foundation (USA). \end{acknowledgments}

\bibliography{auto_generated}
\clearpage
\appendix
\section{Matrix Element Method}
\label{sec:appendix}

As mentioned in Section~\ref{sec:discriminatingObservables}, discriminants based on the MEM approach have been developed for the $2\Plepton\ss+1\tauh$ and $3\Plepton$ categories.
Additional details on their computation are given in this Appendix.
The matrix element (ME) $\mathcal{M}_\Omega(\mathbf{x})$ associated with a given process $\Omega$
depends on a set of kinematic variables $\mathbf{x}$ that corresponds to the four-momenta, at parton level,
of the particles in the initial and final state.
We use bold letters to indicate vector quantities.
The square of the ME is convoluted with a function $W(\mathbf{y} \vert \mathbf{x})$,
referred to as the ``transfer function'' (TF), which represents the experimental resolution.
More specifically, the function $W(\mathbf{y} \vert \mathbf{x})$ corresponds to the probability for measuring a set of observables $\mathbf{y}$ in the detector,
given that the corresponding parton-level momenta are equal to $\mathbf{x}$.

The MEM computes the differential cross section of the process $\Omega$ with respect to the observables $\mathbf{y}$,
while integrating over the unmeasured or poorly measured parton-level quantities $\mathbf{x}$,
as well as over the Bjorken scaling variables~\cite{Bjorkenx} $x_a$ and $x_b$ of the colliding protons.
For each event a weight $w_\Omega(\textbf{y})$ is computed, which quantifies the compatibility of the event, characterized by the measured observables $\mathbf{y}$,
with the hypotheses that the event is produced by the process $\Omega$:
\begin{equation}
w_\Omega(\textbf{y}) \propto \sum\limits_{p} \int{\rd\textbf{x}\,  \rd x_a\, \rd x_b \frac{f_i(x_a,Q)f_j(x_b,Q)}{x_a x_b s} \delta^4(x_a P_a + x_b P_b - \sum p_k) \vert\mathcal{M}_\Omega(\textbf{x})\vert^2 W(\textbf{y} \vert \textbf{x})} .
\label{eq:MEM_weight}
\end{equation}
The sum $\sum\limits_{p}$ extends over all possible associations between parton-level and reconstructed objects.
The square of the ME, $\vert\mathcal{M}_\Omega(\textbf{x})\vert^2$, is computed at LO using the \MGvATNLO program.
The symbols $f_i(x_a,Q)$ and $f_j(x_b,Q)$ denote the PDFs,
which we evaluate numerically using the \textsc{CTEQ6.6}~\cite{CTEQ66} and {NNPDF3.0} LO sets.
The four-dimensional $\delta$-function $\delta^4(x_a P_a + x_b P_b - \sum p_k)$ ensures the conservation of energy and momentum.
The integral on the right-hand side is first transformed analytically,
in order to eliminate the $\delta$-function and to make the computation of the integral numerically tractable,
and then computed numerically using the \textsc{VEGAS} algorithm~\cite{VEGAS}.
A complication arises from the fact that we use LO ME for the $\ttH$ signal and for background processes.
The LO ME strictly applies only to events in which no additional jets, besides the jets corresponding to quarks in the LO ME, are produced in the hard scattering interaction.
At the center-of-mass energies of the LHC the phase space for quantum chromodynamics radiation is large, however,
and particles with masses up to a few hundred \GeV are typically produced in association with a sizable hadronic activity~\cite{Alwall:2010cq}.
In order to use the LO ME, we transform the system of all particles that are present in the LO ME
into a frame in which this system has zero \pt.
In the opposite case of events where the reconstructed jet multiplicity is lower than the number of quarks present in the LO ME
the integral on the right-hand-side of Eq.~(\ref{eq:MEM_weight})
is extended by an integration over the variables associated to the missing jets.
The TFs $W(\mathbf{y} \vert \mathbf{x})$ are obtained from the MC simulation and are used to model
the \pt resolution of jets and the resolution on $\ptvecmiss$.
Separate TFs are used for $\Pbottom$ quark and for light quark and gluon jets.
The TFs are also used to account for the energy loss due to neutrinos produced in the decays of $\Pbottom$ quarks to leptons and in $\Pgt$ lepton decays.
The fraction of $\Pgt$ lepton energy carried by neutrinos is on average higher in $\Pgt\to\Plepton\APnu_{\Plepton}\Pnut$ decays compared to $\Pgt\to\tauh\Pnut$ decays
and separate TFs are determined for both cases.

According to the Neyman lemma~\cite{NeymanLemma}, the ratio of weights $w_\Omega(\textbf{y})$ computed for the signal and for the background hypothesis
constitutes an optimal observable to separate the $\ttH$ signal from backgrounds:
\begin{equation}
\LR(\textbf{y}) = \frac{w_{\ttH}(\textbf{y})}{w_{\ttH}(\textbf{y})+\sum\limits_{\mathrm{B}} \kappa_{\mathrm{B}} \, w_{\mathrm{B}}(\textbf{y})} .
\label{eq:memLikelihoodRatio}
\end{equation}
The coefficients $\kappa_{\mathrm{B}}$ that quantify the relative importance of different background processes $\mathrm{B}$ are determined by a numerical optimization,
in order to achieve the maximal separation of the $\ttH$ signal from all background processes.

\cleardoublepage \section{The CMS Collaboration \label{app:collab}}\begin{sloppypar}\hyphenpenalty=5000\widowpenalty=500\clubpenalty=5000\vskip\cmsinstskip
\textbf{Yerevan~Physics~Institute, Yerevan, Armenia}\\*[0pt]
A.M.~Sirunyan, A.~Tumasyan
\vskip\cmsinstskip
\textbf{Institut~f\"{u}r~Hochenergiephysik, Wien, Austria}\\*[0pt]
W.~Adam, F.~Ambrogi, E.~Asilar, T.~Bergauer, J.~Brandstetter, E.~Brondolin, M.~Dragicevic, J.~Er\"{o}, A.~Escalante~Del~Valle, M.~Flechl, M.~Friedl, R.~Fr\"{u}hwirth\cmsAuthorMark{1}, V.M.~Ghete, J.~Grossmann, J.~Hrubec, M.~Jeitler\cmsAuthorMark{1}, A.~K\"{o}nig, N.~Krammer, I.~Kr\"{a}tschmer, D.~Liko, T.~Madlener, I.~Mikulec, E.~Pree, N.~Rad, H.~Rohringer, J.~Schieck\cmsAuthorMark{1}, R.~Sch\"{o}fbeck, M.~Spanring, D.~Spitzbart, A.~Taurok, W.~Waltenberger, J.~Wittmann, C.-E.~Wulz\cmsAuthorMark{1}, M.~Zarucki
% ...
\vskip\cmsinstskip
% ...
1:~Also at~Vienna~University~of~Technology, Vienna, Austria\\
% ...
\end{sloppypar}
\end{document}